\documentclass[12pt]{article}
\usepackage[utf8]{inputenc}
\usepackage[english]{babel}
\usepackage{graphicx}
\usepackage{tikz}
\usepackage{bbm}
\usepackage{mathtools}
\usepackage{amssymb}
\usepackage{adjustbox}
\usepackage{hyperref}
\usepackage{lscape}
\usepackage{amsmath}
\usepackage{setspace}
\usepackage{color}
\usepackage{booktabs}
\usepackage{longtable}
\usepackage{comment}
\usepackage{multicol}
\usepackage{multirow}
\usepackage{colortbl}
\usepackage{float}
\usepackage{cite}
\usepackage[titletoc]{appendix}
\usepackage{lipsum}
\usepackage[margin=1in,includefoot]{geometry}
\usepackage{fancyhdr}
\usepackage{tabularx}
\usepackage{setspace}

\usepackage{subcaption}
\usepackage{xr}
\numberwithin{equation}{section}

\newtheorem{proposition}{Proposition}

\newtheorem{lemma}{Lemma}

\usepackage{setspace}
\usepackage[round,authoryear]{natbib}
\usepackage{authblk}

\begin{document}

\title{Composite Likelihood for Stochastic Migration Model with Unobserved Factor\thanks{The authors gratefully acknowledge financial support of the chair of the Authority of Prudential Control and Resolution (ACPR)/Risk Foundation: Regulation and Systemic Risks, the ECR DYSMOIA and the Natural Sciences and Engineering Council of Canada (NSERC). We thank the anonymous referee for helpful comments.}}
\author[1]{ Djogbenou, A.}
\author[2]{Gouri\'eroux, C.}
\author[1]{Jasiak J.}
\author[3]{M. Bandehali}
\affil[1]{York University} \affil[2]{Toronto University, Toulouse School of Economics and CREST}\affil[3]{Equitable (EQ) Bank, Toronto} 
\renewcommand\Authfont{\large}
\renewcommand\Affilfont{\small}
\date{ \large{\today}}
\maketitle
\thispagestyle{empty}
\begin{abstract}
\noindent  We introduce the conditional Maximum Composite Likelihood (MCL) estimation method for the stochastic factor ordered Probit model of credit rating transitions of firms. This model is recommended for internal credit risk assessment procedures in banks and financial institutions under the Basel III regulations. Its exact likelihood function involves a high-dimensional integral, which can be approximated numerically before maximization. However, the estimated migration risk and required capital tend to be sensitive to the quality of this approximation, potentially leading to statistical regulatory arbitrage.
The proposed conditional MCL estimator circumvents this problem and maximizes the composite log-likelihood of the factor ordered Probit model. We present three conditional MCL estimators of different complexity and examine their consistency and asymptotic normality when $n$ and $T$ tend to infinity. The performance of these estimators at finite $T$ is examined and compared with a granularity-based approach in a simulation study. The use of the MCL estimator is also illustrated in an empirical application.  \\

\end{abstract}
\newpage

\setcounter{page}{1}
\section{Introduction}

Under the “internal-ratings-based” (IRB) approach advocated in the Basel II and III regulation, banks use their internal risk rating systems  to estimate the risk exposures, credit rating migration probabilities and the probability of default (PD) in order to evaluate their regulatory capital requirements [see  \citet{bas2,bas3}, \citet{hull2012risk}, \cite{GG2016}]. Under Pillar II, financial institutions must also conduct stress tests to determine the level of capital needed to absorb losses in worsening economic conditions and be protected against systemic risk. For these reasons, banks perform their own credit rating migration analysis in order to monitor the changes in borrowers’ credit quality and to predict borrowers’ potential default in a volatile economic environment.
This analysis concerns the ``internal'' or ``in-house'' established credit rating histories of borrowers, classified into credit quality categories, which are determined independently of the ratings publicly provided by the rating agencies such as the Moody's.\footnote{Publicly available credit ratings of large obligors are available from the rating agencies such as the Moody's, Standard and Poor's (S\&P), and Fitch.}     
The internal credit rating analysis is applied to the historical probabilities of default and migration probabilities. It differs from the analysis  of their risk-neutral counterparts, which underlies the pricing of credit derivatives, such as credit default swaps (CDS),  Collaterized Debt Obligations (CDO), or derivatives written on iTraxx [see, e.g. \citet{DEHS2009}, \citet{AGS2018} in continuous time, \citet{GMP2006} in discrete time, \citet{GMMR2021} for joint historical and risk-neutral analysis].  The internal ratings are used for pricing the portfolios of credits offered to a large number of small and medium-size firms whose assets are not traded on the markets. Even for large firms, the historical and risk-neutral probabilities of default can differ significantly. In our paper, the analysis of internal ratings is consistent with prudential banking supervision and aims at avoiding a pure mark-to-market pricing of risk.\footnote{There is often a confusion about the  notions of historical and risk-neutral risks. For example, Moody's Analytics provides ``EDF'' estimates of the historical probability of default by considering default frequencies of firms with the same  distance-to-default (DD). However, the notion of DD is risk-neutral.}

The credit rating migration analysis concerns the changes [i.e. upgrades or downgrades] of borrowers' credit quality over 
time with respect to their previous ratings [\cite{altman1997credit}]. These data are available from  monthly or 
quarterly time series of credit migration matrices comprising the qualitative ratings of firms, ranked from the low risk category 
A to the most risky rating D of default. The ordered Probit model for credit ratings arises as a natural specification, which 
has been extended to the Asymptotic Single Risk Factor (ASRF) model by \citet{Vasicek1991} [see also 
\citet{Vasicek1987}, \cite{NPV2001}]. The ASFR  is a 
stochastic factor probit model of default with an independent and identically distributed common random unobserved 
factor capturing the systemic risk effect. The factor is assumed to drive the parameters of a latent quantitative score 
function in the model, which is transformed into qualitative ratings. Due to the presence of the unobserved 
common factor,  the observed rating histories are  cross-sectionally dependent, which can explain default correlation.
\cite{GG2005a}, \cite{Feng2008},  \ extended this setup to  multiple credit 
rating categories with common systemic factors that can be serially correlated, in order to predict the future credit ratings of firms. This extension is strongly recommended under the Basel III regulatory measures: ``Interdependence between issuers is frequently modelled in a similar way to the regulatory framework, using a combination of an idiosyncratic (i.e. individual) and one or more systemic risk factors'' [\cite{EBA2012}, article 12 on Systemic Risk Factor].
Moreover, the dynamic ordered probit model takes into account the heterogeneity of issuers and satisfies the requirement that ``Separate transition matrices may be applied for specific groups of issuers and geographical areas'' [\cite{EBA2012}]. It also reproduces other stylized facts such as the rating momentum [see e.g. \cite{AK1992}].

The estimation of the ordered Probit model with a latent common factor is challenging.
In order to derive the joint density of observed ratings, the history of the latent factor has to be integrated out. Therefore, the exact likelihood function based on the joint density of rating histories involves an integral of high dimension, increasing with the number of observations over time. Due to the presence of the multiple integrals, the exact maximum likelihood needs to be replaced by an approximation in practice. This paper introduces the conditional Maximum Composite Likelihood (MCL) as an alternative estimation method for the stochastic factor ordered Probit model.
The MCL estimators have been widely used in the statistical literature to handle complex likelihood functions [see, \citet{Lindsay1988}, \citet{Varian2008}, \citet{VRF2011}, \cite{GM2018}].  The conditional composite likelihood functions are obtained by multiplying a collection of conditional component likelihoods, each depending 
on some integrals. In the one factor framework, these integrals are of dimension 1.

There exist alternative approximation methods, most of which involve a set of arbitrary control parameters, having a significant impact on the associated required capital. These parameters are, for example, the discretization steps [\citet{Farmer2021}], tuning parameters, penalties, etc. The effect of the statistical approximation and optimization method can go as far as to partly eliminate the need for keeping an internal capital reserve, which is called a ``statistical regulatory arbitrage''. Therefore  these approximations are often not validated by the supervisory authorities who are regularly auditing the internal databases and estimation techniques.\footnote{``Any estimation technique should be duly justified and documented'' [\cite{EBA2012}].}

So far, the banking supervisory authority has validated selected standardized approximation methods, such as the granularity adjusted approach, that is valid and efficient if both the cross-sectional and temporal dimensions are large [see \citet{GG2014,GG2015}, for general discussion] and the Simulated Maximum Likelihood (SML) method with a large number of simulations [\cite{Feng2008}].
Both these methods circumvent the high-dimensional integration. Under the SML estimation employed in \cite{Feng2008}, the integral is approximated by simulations, allowing for the latent factor values to be filtered out ex-post. The quality of the simulation-based approximation depends on the number of simulations, which can become high, depending on the number of time units considered and the complexity of factor dynamics.
This makes this method computationally intense. The granularity-based approach [\citet{GG2015}] is
a two-step estimation method that eliminates the burden of simulations and provides the estimates of the unknown parameters and unobserved factor values. However, the granularity-based estimator depends in the first step on a set of ``nuisance'' parameters of size $T-1$, which increases the computational complexity of this method.
We show that the proposed MCL estimators are computationally less intense than the granularity-based approach, and are reliable in finite sample.

In the panel analysis of credit ratings,  the number of firms determines the cross-sectional dimension $n$, and the number of observed time units determines the  dimension $T$.  Two different asymptotics are considered, when $n,T$ both tend to infinity, or $n \rightarrow \infty$ and $T$ fixed, i.e. finite sample in $T$. In practice, $n$ is often large, while $T$ can be rather small. Therefore, these two types of asymptotics are compared.
When both $n$ and $T$ tend to infinity, the new conditional composite maximum likelihood estimators are shown to be consistent, but not fully efficient, while the granularity-based estimator is consistent and asymptotically efficient. However, when $n$ is large and $T$ is fixed, all estimators converge to stochastic limits that depend on the latent factor values and differ from the true values of the parameters. 

\indent This paper is organized as follows. Section 2 compares the credit rating models that exist in the literature and describes the ordered probit model of credit rating transitions. 
 Section 3 introduces the conditional composite maximum likelihood estimators and the granularity approach. The order and rank conditions for identification are also provided.  Section 4 derives 
the asymptotic properties, i.e. the consistency, rates of convergence and asymptotic normality when both $n$ and $T$ tend to infinity, and also when $n$ tends to infinity and $T$ is fixed. In Section 5, the performance of MCL and granularity-based estimators in finite sample is examined in a simulation study. Section 6 includes the empirical application. The observed transition probabilities are computed from the Compustat Standard and Poor's (S\&P) rating database from 1985Q4 to 2016Q4, available through Wharton Research Data Services. We analyze the estimated parameters, probabilities of defaults, and the downgrade probabilities at different horizons. 
Section 7 concludes the paper. Proofs are given in Appendices A-C and the simulation details and additional simulation results are presented in the online Appendix D. Throughout the paper, variables indexed by $t$, $i$ in parenthesis denote random sequences.

\section{The Stochastic Factor Ordered-Probit Model}

In this section, we focus on the stochastic factor ordered probit model and its state space representation.
The expression of the complete likelihood function is derived, highlighting the presence of multiple integrals of large dimension. Next,
we discuss the models of joint evolution of individual ratings that already exist in the literature. 

Let $y_{i,t}$ denote the rating of firms $i, \; i=1,...,n$ at time  $t, \; t=1,...,T$. The ratings are qualitative variables that take $K$ values associated with different rating categories. The sequences of variables $y_{i,t} \; t=1,...,T $  for $i=1,...,n$ represent the panel of qualitative individual histories of credit ratings. The migration model defines the joint distribution of the qualitative variables $y_{i,t}, \; i=1,...,n, \; t=1,...,T$ and provides information on the transitions (migrations) of individuals (firms) between the ratings.

Let us consider the following extention of the standard Vasicek model of default risk [\cite{Vasicek1991}, \cite{GL2013}, \cite{GG2016}] to a migration model with a latent factor. The unobserved (latent) stochastic common factor is denoted by $f_t$. 

We assume that the conditional transition probability of the factor given the whole past information depends on $f_{t-1}$ only , i.e.,

$$l(f_t|f_{t-1}, f_{t-2},... ;\;y_{i, t-1}, y_{i, t-2},...; i=1, ...,n) = l(f_t |f_{t-1}).$$

In addition, we assume conditional on the path of the common factor, the individual rating histories are independent, heterogeneous
Markov chains with

$$P[y_{i,t} = k| y_{i,t-1} = j, f_t] = p_{jk}(f_t).$$

\noindent Under the above assumptions, the joint process $(y_{i,t}, i=1,...,n, f_t)$ is a Markov process with an exogenous evolution of factor process $(f_t)$.

Since the factor is unobserved, its evolution has to be integrated out to get the joint distribution of individual histories. This creates migration (and default) correlation because the factor is common to all individuals (firms). The interdependence of risks has to be included as an incremental risk, accounted for by additional required capital [see \cite{bas3},  \cite{EBA2012}].
It also implies non-Markovian features after integrating  the conditional transition probability  
with respect to factor $f = (f_t)$. In particular, the conditional transition probability

\noindent $P[y_{i,t+1} = k, y_{i,t+2} = k| y_{i,t} = j] $, for example, is not equal to 
$P[y_{i,t+2} = k| y_{i,t+1} = k] P [y_{i,t+1} = k| y_{i,t} = j] $.

\noindent Indeed, we have
$$P[y_{i,t+1} = k, y_{i,t+2} = k| y_{i,t} = j] = \frac{ E P[y_{i,t} = k, y_{i,t+1} = k, y_{i,t+2} = j|f] } {E P[y_{i,t} = j| f]},$$
   
\noindent where the expectation is taken with respect to the stochastic evolution of $f$ over the period $(0,t+2)$, which has a different impact on the probability of staying in state $k$ depending on the last transition being an up- or down-grade, and the date of that transition.

Therefore, the model account for the rating momentum effect, i.e. the fact that the intensity of transitions out of 
a given state is influenced by previous transitions into that state, and more generally for non-Markovian features [\cite{GK2009}].

This unobserved factor model can be viewed as an infinite mixture model at time $t$, with stochastic weights.  An example of this type of model is the stochastic factor ordered probit model [\cite{GG2005a, GG2014}, \cite{Feng2008}, \cite{HZ2015}, \cite{CLP2021}] examined in this paper.\footnote{or its continuous time counterparts, i.e. the dynamic marked point processes with common systemic factors (see \cite{Creal2013}, Section 4.3, \cite{Koopman2008} for a continuous time approach without systemic factor).} Its state-space representation is given below.  
 
\subsection{The State-Space Representation}
\indent Let $y_{i,t}^*$ and  $y_{i,t}$ denote the (credit) score and rating of  firm $i$, $ i = 1, . . . , N$ at time $t$,  $t = 1, \ldots, T$.  The latent continuous quantitative score $(y^{*}_{i,t})$ determines the individual qualitative rating $y_{i,t}$. More precisely, the
quantitative score is discretized in order to obtain the individual qualitative ratings. Therefore, an observed rating is determined as follows,
\begin{equation}
y_{i,t}=k, \textrm { if and only if } c_{k} \leq y_{i,t}^{*} < c_{k+1},\;\; k=1,...,K,
\end{equation}   
where $c_1< \cdots <c_{K+1}$ are the thresholds. Relation (2.1) shows how the observable endogenous credit rating $(y_{i,t})$ is linked to the latent score function $(y_{i,t}^{*})$. By convention, we have $c_{1}=-\infty$ and $c_{K+1}= +\infty$. Relation (2.1) defines the measurement equation of the state space representation of the model.

The conditional distribution of the quantitative scores given the factor path and the previous scores depends on the common latent factor  $f_t$ 
and on the past individual ratings $y_{i,t-1}$, such that
\begin{equation}
y_{i,t}^{*}= \delta_{j}+\beta_{j}f_{t}+\sigma_{j}u_{i,t}, \;, i=1,...,n, \; \textrm{ if }  y_{i,t-1}=j , j=1, ..., K, t = 2, \ldots, T,
\end{equation}
and $y_{i,1}$ is the first observed rating for firm $i$. The multivariate, continuous, latent processes $y^{*}_{i,t}$, are generated by individual level effects $(\delta_{j})$, volatility effects $(\sigma_{j}), \sigma_j>0$, factor effects where the components of $\beta_{j}$ define the factor sensitivities. When coefficient $\beta$ is large (small, resp.), the effect of systemic risk carried through the factor is strong (weak, resp.). 
All the parameters $\delta_j$, $\beta_j$, $\sigma_j$ depend on the previous rating $j$.
While the idiosyncratic risks $(u_{i,t})$ can be diversified, the systemic risk $(f_{t})$ cannot be diversified. Thus the presence of systemic risk generates risk interdependence in the model. Because parameters $\beta$ are different in each rating category, the risk interdependence varies across rating transitions resulting in risk momentum. Among these parameters, $\delta_j$ and $\sigma_j$ summarize the effect of idiosyncratic risk, and $\beta_j$ is the sensitivity to systemic risk.

The following autoregressive model of order 1 (AR(1)) represents  the common factor dynamics,
\begin{equation}
f_{t}=\rho f_{t-1}+\sqrt{1-\rho^{2}}\eta_{t}, |\rho|<1,  t = 2, \ldots, T,
\end{equation}
where $\eta_t$ defines the shock to the common factor and $f_1$ is drawn in the stationary distribution. Alternatively, a multidimensional factor can be considered to distinguish between the dynamic migration patterns of firms with good and poor credit quality, respectively. See, for example, \citet{GG2014}, for more details.

The system of equations (2.2)-(2.3) defines the state equations of the state-space model. Let us introduce the following assumptions to obtain a migration model with migration correlation and rating momentum.

\noindent {\bf Assumption A.1:} The errors $u_{i,t}, \; \eta_t, \; i=1,...,n, t=1,...,T$, are independent, standard normal variables.

The independence assumption allows for performing impulse response analysis by shocking separately the idiosyncratic and systematic innovations, to perform a stress-test in particular. The assumption of identical distribution and the fact that coefficients 
in (2.2) are independent of the firm implies that we consider a homogeneous set of firms, obtained by crossing the country, industrial sector and firm  size, in compliance with the current regulation. 


\noindent {\bf Assumption A.2:} The factor process $(f_t)$ is the strongly stationary solution of autoregressive equation (2.3).


As the processes ($f_t$), ($u_{i,t}$), $i=1,\ldots,n$, are independent and strictly stationary, it follows that the joint $n-$dimensional process $y_t^\ast=\left(y_{1,t}^\ast,\ldots,y_{n,t}^\ast\right)^\prime$ is also strictly stationary, and so is its state discretized version $y_t=\left(y_{1,t},\ldots,y_{n,t}\right)^\prime$. However, the individual components ($y^\ast_{i,t}$), $i=1,\ldots,n$ are not independent due to the effect of the common factor $f_t$.\footnote{In this respect this model differs from \citet{Tuzcuoglu2019}, 
where the state equations (2.2)-(2.3) are replaced by $ y_{i,t}^* = \rho y_{i,t-1}^* + \beta' x_{i,t} + \alpha_i + \epsilon_{i,t},$
with independent 
$(\alpha_i, (\epsilon_{i,t})),\; i=1,...,n,$ given $x$. This specification does not contain systemic risk and does not allow for risk interdependence.}

It is important to notice that the error variance in equation (2.3) has been set equal to $1-\rho^2$. This implies that factor $f_t$ is marginally normally distributed with mean 0 and variance 1: $E(f_t)=0, Var(f_t) =1$. These moment restrictions are introduced to solve the factor identification issue, since the factor is defined up to a linear affine transformation. 

 In practice, the underlying quantitative scores are computed by a credit institution and each individual (firm) can request the records of its own score history. However, the complete score database is, in general, proprietary and the information on the quantitative scores is not available to an outsider econometrician/data scientist. 
The factor $f_t$ is assumed unobserved for the following two reasons: First it creates the cross-sectional correlation between individual risks. Second, it provides a  dynamic model that can be used to predict the future defaults. A bias could result from directly replacing factor $f_t$ by an observed proxy $\hat{f}_t$,
such as the VIX market volatility index, a consumer sentiment index, consumption growth, a business cycle indicator [see e.g. \citet{BDDF2018}, \citet{AGS2018}], or the slope of the yield curve. Moreover, if factors are observed,  their predictions cannot be computed without specifying an additional model of the dynamics for all the observed factors in $\hat{f}_t$, and checking that these observed factors are exogenous. 

\subsection{The Complete Likelihood Function}
In order to derive the joint density of observations $y_{i,t}, \; i=1,...,n, \; t=1,...,T$, the unobserved factor path has to be integrated out. As a consequence, observations $y_{i,t}$ are cross-sectionally dependent and serially dependent with a non-Markovian serial dependence. More precisely, the stochastic migration probabilities between dates $t-1$ and $t$, conditional on $f_t$, are given by
\begin{align}
p_{jk,t}&=p_{jk}(f_{t};\theta)=
P[y_{i,t}=k|y_{i,t-1}=j, f_{t}]  \nonumber \\
&= P[c_{k} \leq y^{*}_{i,t}<c_{k+1}|y_{i,t-1}=j, f_{t}]  \nonumber \\
&= \Phi \left( \frac{c_{k+1}-\beta_{j}f_{t}-\delta_{j}}{\sigma_{j}}\right) - \Phi\left(\frac{c_{k}-\beta_{j}f_{t}-\delta_{j}}{\sigma_{j}}\right), j,k=1, ..., K,\; t=2,...,T,
\end{align}
where $\Phi$ denotes the cumulative distribution function (c.d.f.) of the standard normal. Thus each row of the transition matrix conditional on $(f_t)$ contains an ordered polytomous probit model with a common explanatory factor $f_{t}$. When factor $f_{t}$ is unobserved stochastic and serially correlated as in (2.3), the transition matrices are stochastic and serially dependent. 
\medskip

Let us now define the log-likelihood function of the stochastic migration model. The vector $\theta$ includes the parameters of the state space model, which are parameters $\delta_j,\beta_j, \sigma_j,j=1,\ldots,K$ in the quantitative score, and parameters $c_k, k=2,\ldots,K$ defining the states. As the conditional migration matrices are functions of parameter vector $\theta$ as well as of the common factor values $f=(f_{t})$, the likelihood function conditional on $f$ and the initial rating $y_1$ is
\begin{equation}
L_T(Y|f,y_1;\theta)=\prod_{t=2}^{T}\prod_{k=1}^{K}\prod_{j=1}^{K} (p_{jk}(f_{t};\theta))^{n_{jk,t}},
\end{equation}
\noindent where $n_{jk,t}$ denotes the number of firms which migrate from $j$ to $k$ between $t-1$ and $t$, $Y=(y_{i,t})$ for $i=1, ..., n$ and $t=2, ..., T$, $f=(f_t,\; t=2,\ldots,T)$, and $y_1=\left(y_{1,1},\ldots,y_{n,1}\right)^\prime$.

Since the factor history is not observed, after integrating out the factor values $(f_{2}, ..., f_{T})$, the log-likelihood function, given the initial value $y_1$ only, is
\begin{equation}
\ell(Y|y_1;\theta, \rho)=log \int ... \int\prod_{t=2}^{T}\prod_{k=1}^{K}\prod_{j=1}^{K} \left[(p_{jk}(f_{t} ;\theta))^{n_{jk,t}} \psi(f_{2}, ..., f_{T};\rho) \right] df_{2} ... df_{T} ,
\end{equation}

\noindent where $\psi$ refers to the joint probability distribution function of factor values. The above log-likelihood function contains a multivariate integral. The dimension of this integral is of order $T-1$, as there is a common factor value for each transition at time $t$.  Therefore the exact computation of this likelihood is infeasible and its approximation is often not sufficiently robust.\footnote{See \cite{Feng2008} for the discussion of robustness when simulations are used.} The MCL estimators are  convenient alternatives for complicated nonlinear dynamic state-space models allowing for circumventing the high-dimensional integral.

\subsection{Migration Models in the Literature} 
In the literature, there exist alternative models of rating transitions that rely on simplified assumptions. For comparison, these models can be classified as the Markov chain models and models with a rating momentum adjustment.

\subsubsection{Markov Chain}

The basic migration model 
assumes that  the histories of individual transitions (migrations) of firms between the ratings (states)
are independent, identically distributed Markov chains.
The assumption of independent Markov chains implies that the individual histories ($y_{i,t}, \; t=1,...,T$) are independent across individuals (firms) $i$, $i = 1,...,n$. These Markov chains are homogeneous with transition probabilities $P[y_{i,t} = k| y_{i,t-1} =j] = p_{jk}$ independent of the firm.\footnote{We choose the transition matrix $P=[p_{jk}]$ such that the elements of each row sum up to one.} This model is examined in  \cite{LS2002}, Section 3, and reviewed in  \cite{DPS2020}, for example. It does not account for the rating momentum and it does not assume  migration correlations among the individuals, implying no default correlation among the firms.

\subsubsection{Adjustment for rating momentum}

The rating momentum is accounted for in a non-Markovian model, such as the hazard model
introduced in  \cite{LS2002} and \cite{DPS2020}, Section 4, based on the assumption of Semi-Markov chains. This assumption implies that the individual histories are independent. Then, the transition probabilities are given by
$$P[y_{i,t} = k| y_{i,t-1}, y_{i,t-2},,...] = p_{jk} exp( c Z_{i, t-1}),$$
\noindent where $y_{i,t-1}=j$ and $Z_{i,t-1} = 1$, if firm $i$ was downgraded to its current state,  $Z_{i,t-1} = 0$, 
if firm $i$ was upgraded to its current state. 

Alternatively, the non-Markovian feature can be introduced by considering a mixture of Markov chains [\cite{FS2008}].
Under the assumption of Markov Chain mixture, the individual histories are independent and the transition probabilities are obtained from a mixture of Markov chains.

Other non-Markovian assumptions can be introduced by considering time varying exogenous variables, such as the observed regime of business cycle [\cite{BDKSS2002}, \cite{GS2014}], or time itself.
These extensions are characterized by the independence of migrations, implying no migration correlation among the
individuals (firms). Hence, drawback b) is the common issue of all these models.

Other models without a latent factor can be found in \citet{RC2017}, \citet{MT2012}, and  \citet{HVH2022}. Alternatively to the frequentist approach in this paper, Bayesian procedures like the one in \citet{STT2009} could be used if the researcher is willing to impose priors on the different parameters.

\section{Conditional Composite Likelihood for Migration Model with Unobserved AR(1) Factor}
\subsection{Expected Transition Probabilities}

\indent The process of transition matrices $\{P_{t}, t=1, ..., T\}$ has component matrices $P_{t}=\left(p_{jk,t}\right)$, which provide the probabilities of transitions from state $j$ to state $k$ between times $t-1$ and $t$ given $f_t$. From (2.4), it follows that the elements of matrix $P_{t}$ are
$$p_{jk,t}=p_{jk}(f_{t};\theta)= \mathbb{P}[y_{i,t}=k|y_{i,t-1}=j,f_{t}]
=\Phi\left(\frac{c_{k+1}-\beta_{j}f_{t}-\delta_{j}}{\sigma_{j}}\right)-\Phi\left(\frac{c_{k}-\beta_{j}f_{t}-\delta_{j}}{\sigma_{j}}\right),$$
\noindent $\; k, j =1,...,K .$

Let us now compute the product of two successive transition matrices $P_{t}^{(2)}=P_{t}P_{t-1}$ to obtain the probabilities of transition at horizon 2 from state $j$ to $k$ between times $t-2$ and $t$ given the factor history $\left(f_t\right)$. The elements of matrix $P_{t}^{(2)}$ depend on $f_t, f_{t-1}$ and are given by
\begin{equation}
p_{jk,t}^{(2)}=p_{jk}(f_{t},f_{t-1};\theta)=\mathbb{P}[y_{i,t}=k|y_{i,t-2}=j,f_{t},f_{t-1}]
=\sum_{l=1}^{K} [ p_{lk}(f_{t},\theta) p_{jl}(f_{t-1},\theta)].
\end{equation}
They can be computed from the elements of matrices $P_{t}$ and $P_{t-1}$. Let us denote by $P$ and $P^{(2)}$ the expectations of matrices $P_{t}$ and $P_{t}^{(2)}$ with respect to the common factor history\textbf{, i.e,}
\begin{equation}
 P=E(P_{t}), \;\;\; P^{(2)}=E(P_{t}^{(2)})=E(P_{t}P_{t-1}).
\end{equation}
The elements of matrix 
$$ P=[p_{jk}]=[p_{jk}(\theta)]=E_{f_{t}}[p_{jk}(f_{t},\theta)],$$
are obtained by integrating out the unobserved factor value $f_t$.
\begin{lemma}
Under Assumptions A1 and A2, we have
\begin{align*}
 p_{jk}(\theta) & =\Phi\left(\frac{c_{k+1}-\delta_{j}}{\sqrt{\sigma_{j}^{2}+\beta_{j}^{2}}}\right)-\Phi\left(\frac{c_{k}-\delta_{j}}{\sqrt{\sigma_{j}^{2}+\beta_{j}^{2}}}\right).
\end{align*}
\end{lemma}
\indent \textbf{Proof}. See Appendix A.1.

It is easy to see that the expected matrix $P$ is also a transition matrix. Indeed, its elements are non-negative and additionally $Pe = E(P_t)e = E (P_t e) = e$, where $e$ is a vector of ones. Matrix $P$ is a quasi-transition matrix for $y_{i,t}$, since $p_{jk}$ is not equal to the conditional probability of $y_{i,t }=k$ given $y_{i,t-1}=j$, except
when the $f_t$'s are i.i.d., i.e., when $\rho=0$. In fact, it is computed as if $f_t$ was independent of $y_{i,t-1}$. Each row of this quasi-transition matrix corresponds to another ordered probit model.

\noindent The elements of matrix $P(2)$ are obtained by integrating jointly with respect to $f_t, f_{t-1}$. We have
\begin{align*}
   P^{(2)} & =[p_{jk}^{(2)}]=[p_{jk}^{(2)}(\theta,\rho)]=E_{f_{t}, f_{t-1}}\left[\sum_{l=1}^{K}p_{lk}(f_{t},\theta)p_{jl}(f_{t-1},\theta)\right].
\end{align*}
\begin{lemma}
Under Assumptions A1 and A2, we have
$$p_{jk}^{(2)}(\theta,\rho)=\int \sum_{l=1}^{K} \Bigg[\bigg[\Phi\left(\frac{c_{k+1}-\delta_{l}-\beta_{l}\rho f}{\sqrt{\sigma_{l}^{2}+\beta_{l}^{2}(1-\rho^{2})}}\right)-\Phi\left(\frac{c_{k}-\delta_{l}-\beta_{l}\rho f}{\sqrt{\sigma_{l}^{2}+\beta_{l}^{2}(1-\rho^{2})}}\right)\bigg]$$ $$\times\bigg[\Phi\left(\frac{c_{l+1}-\delta_{j}-\beta_{j}f}{\sigma_{j}}\right)-\Phi\left(\frac{c_{l}-\delta_{j}-\beta_{j}f}{\sigma_{j}}\right)\bigg]\Bigg]\\ \phi(f)df,$$
where $\phi$ is the probability distribution function (pdf) of the standard normal.
\end{lemma}
\textbf{Proof}. See Appendix A.2. 

$P^{(2)}(\theta, \rho)$ is a quasi-transition matrix at horizon 2 computed as if $(f_t, f_{t-1})$ were independent of $y_{i, t-1}$.
The quasi-transitions at horizon 2 involve one-dimensional integrals only, which are easy to compute numerically.

\subsection{Conditional Composite Likelihood Functions}

\indent This section presents the conditional composite likelihood functions for the migration model with an unobserved AR(1) factor. The composite likelihoods are often based on misspecified likelihoods, which are easier to calculate [see \citet{CR2004}, \citet{VRF2011}]. In our framework, the conditional composite likelihoods are constructed from the quasi-migration probabilities at horizons 1 and 2 to reduce the dimension of the integrals. We also present the conditional likelihood used in the first step of the granularity approach.

As mentioned earlier, the parameter vector $\theta=(\beta_k, \delta_k, \sigma_k, c_k)$ includes the parameters characterizing the latent quantitative score, representing the systemic and idiosyncratic risks, and the thresholds that define the qualitative rating category associated to the latent quantitative score. The additional parameter $\rho$ allows for predicting the future systemic risk. Some among the estimation methods given in this section are focused on the rating parameters $\theta$, while others concern both $\theta$ and serial dependence parameter $\rho$.

i) \textbf{The Conditional Composite Log-Likelihood at Lag 1}

\indent The conditional composite log-likelihood function at lag 1, called CL(1), is focused on parameter $\theta$. The associated log-likelihood $L_{cc}(\theta)$ is defined as
\begin{equation}
L_{cc}(\theta)
= \sum_{t=2}^{T} \sum_{k=1}^{K}\sum_{j=1}^{K} \left[\pi_j \hat{p}_{jk,t} \ log( p_{jk}(\theta))\right],
 \end{equation}
\noindent where $\hat{p}_{jk,t}= n_{jk,t}/n_{j, t-1}$ is the observed transition frequency  from $j$ to $k$ in one step over the period $(t-1,t)$, $n_{j, t-1}$ is the count of firms with rating $j$ at the beginning of period $t$, and  $\pi_j, \; j=1,...,K$ is a given set of weights. The log-likelihood $L_{cc}$ is calculated as if the observed ratings 
$(y_{i,t})$, $i=1, ..., n$, were independent across the individuals, while in reality they are linked by the common factor. Moreover, $L_{cc}$ considers the rating processes $(y_{i,t})$, $i=1, ..., n$, as if  these were components of a Markov chain with quasi-transition matrix $P$, although $(y_{i,t})$, $i=1, ..., n$, are not Markov because integrating the factor out increases the memory of the process.
 It also assumes a time independent rating structure $(\pi_j, \; j=1,...,K)$. Therefore, the CL(1) is a quasi (pseudo) log-likelihood. The conditional composite log-likelihood CL(1) depends on parameter vector $\theta$ only, and cannot be used to estimate the factor dynamics, i.e, the autoregressive coefficient $\rho$. For that purpose, it is necessary to increase the lag. 
    
ii) \textbf{The Conditional Composite Log-Likelihood at Lag (2)}

\indent The conditional composite log-likelihood at lag (2), called CL(2), depends on both parameters $\theta$ and $\rho$. The log-likelihood $L_{cc,2}(\theta,\rho)$ is given by

\begin{equation}
L_{cc,2}(\theta,\rho)
= \sum_{t=3}^{T} \sum_{k=1}^{K}\sum_{j=1}^{K} \left[ \pi_j \hat{p}_{jk,t}^{(2)} \ log \ p_{jk}^{(2)} (\theta,\rho)\right], 
\end{equation}

\noindent where $\hat{p}_{jk,t}^{(2)}$ is the observed transition frequency from state $j$ to $k$ in two steps over the period $(t-2,t)$ and $\pi=(\pi_j,\; j=1,...,K)$ is a fixed structure of ratings.

\noindent The composite log-likelihood function $L_{cc,2}(\theta,\rho)$ is computed from the density of $(y_{i,t})$  conditional on $(y_{i,t-2})$ as if the rating histories $(y_{i,t})$ were cross-sectionally independent from one another, $(y_{i,t-2})$ were containing all information about the past and were based
on quasi-transitions over 2 steps. Therefore, the CL(2) is a quasi (pseudo) log-likelihood too.

An important difference between $L_{cc}$ and $L_{cc,2}$ is the set of identifiable parameters. As mentioned above, we can expect to identify $\theta$ from $L_{cc}$, but we cannot identify parameter $\rho$  characterizing the cross-sectional dependence. $L_{cc,2}$ provides additional information that is sufficient to identify $\rho$.

iii) \textbf{The Conditional Composite Likelihood up to Lag 2}

\indent The conditional composite log-likelihood up to lag 2, CL(1,2), is defined  by summing up the previous composite log-likelihoods at lags 1 and 2,
\begin{equation}
L_{c}(\theta, \rho) = L_{cc,2}(\theta,\rho) + a L_{cc}(\theta). 
\end{equation}

\noindent where $a$ is a constant to be selected.\footnote{$a$ could be optimally selected to increase the efficiency of the estimator in a two step approach [Cox, \cite{CR2004}, p.730].} It concerns both parameters $\theta$ and $\rho$.
This  objective function cannot be interpreted as a quasi-likelihood.

iv) \textbf{The Granularity-Based Conditional Log-likelihood }

Let us now introduce another type of log-likelihood for the estimation of parameter $\theta$.
\noindent As shown in Section 2.2, the complete log-likelihood has a complicated expression including a high-dimensional integral of a dimension increasing with $T$. The granularity approach [\cite{GG2005a,GG2015}] replaces the complete log-likelihood by an appropriate expansion for large $T$. This leads to a two step estimation method where, in the first step, the factor values are considered as fixed time effects. 
This log-likelihood conditional on $(f_2,...,f_T)$ is
\begin{align}
L(\theta, f_{2}, ..., f_{T})=&\sum_{i=1}^{n}\sum_{t=2}^{T}\sum_{k=1}^{K}\sum_{j=1}^{K} \left[\mathbbm{1}(y_{i,t}=k,y_{i,t-1}=j) \ log \ p_{jk}(f_{t};\theta)\right]\nonumber\\
&= \sum_{t=2}^{T}\sum_{k=1}^{K}\sum_{j=1}^{K} \left[n_{jk,t} \ log \ p_{jk}(f_{t};\theta)\right] \nonumber\\
&= \sum_{t=2}^{T}\sum_{k=1}^{K}\sum_{j=1}^{K} \left[ n_{j., t-1} \hat{p}_{jk,t} \ log \ p_{jk}(f_{t};\theta) \right],
\end{align}
\noindent where $n_{jk,t}$ (resp. $n_{j,t-1}$) counts all transitions from $j$ to $k$  (resp. is the structure of ratings at $t-1$). 
It is maximized with respect to both parameter $\theta$ and factor path $f_2,...,f_T$ subject to the identification restrictions

\begin{equation}
\frac{1}{T-1} \sum_{t=2}^T f_t = 0,\;\;\;\; \frac{1}{T-1} \sum_{t=2}^T f_t^2 = 1.
\end{equation}

These restrictions on time fixed effects $(f_t)$ approximate the identification restrictions on the latent stochastic factor, i.e. $E (f_t) =0, Var(f_t) =1$.
This conditional constrained log-likelihood 
resembles the composite log-likelihood $L_{cc}$ except that in the composite log-likelihood, $p_{lk}(\theta)$ was made independent of $f_t$ by marginalizing and the observations
are introduced with a fixed rating structure $\pi=(\pi_j, j=1,....,K)$. Since this objective function is maximized with respect to  $\theta, f_2, \ldots ,f_T$, it provides not only an estimator of $\theta$, but also an approximation $\hat{f}_t$ of the factor values. In the second step, an estimator of $\rho$ is obtained by regressing  $\hat{f}_t$ on  $\hat{f}_{t-1},t=2,\ldots,T$. 

Let us focus on parameter $\theta$ and briefly discuss the expected properties of the above estimation methods. 
In the panel framework involving both $n$ and $T$, various notions of asymptotics can be considered.
When $n$ and $T$ both tend to infinity, the granularity approach provides consistent and asymptotically efficient estimators [\cite{GG2014}]. In Section 4, we prove that both CL(1) and CL(2) methods also provide consistent estimators of $\theta$.
In practice, the cross-sectional dimension $n$ is large, but $T$ is much smaller. Therefore, we expect  finite sample effects in $T$ affecting all the estimators. When $n$ tends to infinity, $T$ is fixed, all estimators converge to pseudo-true values $\theta_{\infty}(f_{0,2},...,f_{0,T})$ depending on the latent factor values $f_{0,2},...,f_{0,T}$, including the first step  of the granularity approach due to replacing the factor identification restrictions $E (f_t) =0, Var(f_t) =1$ by their fixed effect counterparts. Hence, for $T$ fixed, all the estimators considered are asymptotically (in $n$) biased both conditionally on factor values and after re-integrating the factor.
Let us now discuss their variances conditional on $f_{0,2},...,f_{0,T}$. In this respect, it is important to consider the number of ``nuisance'' parameters in each of the estimation method: no nuisance parameter in CL(1),
one nuisance parameter $\rho$ in CL(2), $T-1$ nuisance parameters $f_{0,2},...,f_{0,T}$ in the first step of the granularity approach.\footnote{The number of nuisance parameters quickly increases if more systemic risk factors are introduced.} The variances can increase with the number of nuisance parameters.
Moreover, the bias and variance trade-off\footnote{In statistics, the variance-bias trade-off is through the quadratic loss = variance + squared bias. In credit portfolios, it is through a Value-at-Risk of the type VaR= bias + 1.96 $\sqrt{\mbox{variance}}$.} would depend on the dynamic pattern of the true factor values $f_{0,2},...,f_{0,T}$, in particular if they approximately satisfy the restrictions of zero sample mean and unit sample variance for the granularity approach, or are more or less erratic for the CL(1) and CL(2) methods.

\subsection{Identification}

In this section, the order and rank conditions for identification of each of the conditional composite log-likelihoods are discussed. The identification of $\theta, \rho$ in the conditional granularity approach has already been examined in \cite{GG2005a, GG2015}.

\indent The parameters to be identified and their respective numbers are as follows,
\begin{align*}
c_{k}&, \ \ k=2, ..., K, \ \ number: \ \ K-1,\\
\delta_{k}&, \ \ k=1, ..., K, \ \ number: \ K,\\
\beta_{k}&, \ \ k=1, ..., K, \ \ number: \ \ K,\\
\sigma_{k}&,  \ \ k=1, ..., K, \ \ number: \ \ K,\\
\rho&, \ \ number: \ \ 1.
\end{align*}
The total number of independent parameters to identify is $4K-2$. The negative two is due to the score $y_{i,t}^{*}$ being defined up to an increasing function. As we have supposed that it was a linear function of factor $f_{t}$, the score $y_{i,t}^{*}$ is defined up to a linear affine increasing function. The intercept and slope of that linear function are not identifiable. 

\subsubsection{Order Conditions}
\indent In this subsection, the order conditions for each conditional composite log-likelihood are discussed.
These conditions are derived from the probabilities $p_{jk}(\theta), p_{jk}^{(2)} (\theta, \rho)$ that appear in the 
composite log-likelihoods. These probabilities can be consistently estimated if $n$ and $T$ tend to infinity\footnote{They cannot be consistently estimated otherwise, in particular when $n\rightarrow \infty$, $T$ fixed. Indeed, in such a panel framework, the identification does not necessarily imply the existence of a convergent estimator.} (see Section 4). 

i) \textbf{Identification of $\theta$ under CL(1)}

\indent The identifying functions are the reduced form parameters in the CL(1) objective function, i.e. the elements $p_{jk}(\theta)$ of the quasi-transition matrix $P$. There are $K(K-1)$ of these elements that are linearly independent because of the unit mass restriction on each column. Hence, the order condition is
\begin{align*}
    K(K-1)\geqslant 4K-1 \iff  K^{2}-5K+1\geqslant 0,
\end{align*}
by taking into account the absence of parameter $\rho$ in the objective function. This order condition is satisfied for $K\geqslant 5$.

ii) \textbf{Identification of $\theta$  under CL(2)}

\indent The identifying functions are determined by observing that the factor $f$ varies within the integral expression of $p_{jk}^{(2)} (\theta, \rho)$ (see Lemma 2). These identifying functions and their respective numbers are as follows:
\begin{align*}
(1)& \ \ \ \frac{c_{k}-\delta_{j}}{\sqrt{\sigma_{j}^{2}+\beta_{j}^{2}(1-\rho^{2})}}; \ \ \ number: \ \ K(K-1); \\   
(2)& \ \ \ \frac{\epsilon\beta_{j}\rho}{\sqrt{\sigma_{j}^{2}+\beta_{j}^{2}(1-\rho^{2})}}; \ \ \ number: \ \ K; \\
(3)& \ \ \ \frac{c_{k}-\delta_{j}}{\sigma_{j}}; \ \ \ number: \ \ K(K-1); \\
(4)& \ \ \ \frac{\epsilon\beta_{j}}{\sigma_{j}}; \ \ \ number: \ \ K,
\end{align*}
where $\epsilon=\pm 1$ is an unknown sign, since the distribution of $f$ is symmetric. This implies that the integral expression in Lemma 2 is also valid with $f$ replaced by $-f$. There is only one such invariance property and therefore the sign $\epsilon$ is equal for all $j$. The total number of identifying functions of parameters is $2K(K-1)+2K=2K^2$. Hence, the order condition is
\begin{align*}
    2K^2\geqslant 4K-2
  &  \iff    K^{2}-2K+1\geqslant 0\\
   &  \iff  (K-1)^{2}\geqslant 0. 
\end{align*}
The order condition holds for any $K$. 

iii) \textbf{Identification of $\theta, \rho$ under CL(1,2)}

\indent The total number of functions available is equal to the sum of functions available for each component of the total composite log-likelihood. Therefore, the order condition is
\begin{align*}
    3K^2-K\geqslant 4K-2.
\end{align*}
\noindent The order condition is satisfied for any $K$.
\subsubsection{Rank Conditions}

The rank conditions are important for the local identifiability. They are derived for the CL(1) and CL(2) approaches and are similar to the rank condition derived for the granularity approach in \cite{GG2005a} and \cite{GG2015}, p.84. 

\begin{proposition}
Under the CL(1) log-likelihood function and the identifying constraints $c_{2}=0, \gamma_{1}=1$, we can identify the thresholds $c_{k}, \ k=2, ..., K$, the intercepts $\delta_{j}, \ j=1, ..., K$, and the  $\gamma_{j}=\sqrt{\beta_{j}^{2}+\sigma_{j}^{2}}, \ j=2, ..., K$. 
\end{proposition}
\indent \textbf{Proof}. See Appendix B.1.

\begin{proposition}
Under the CL(2) composite log-likelihood function and the identifying constraints $c_{2}=0, \gamma_{1}=\sqrt{\sigma^2_1+\beta^2_1\left(1-\rho^2\right)}=1$, all parameters are identified up to the common sign $\epsilon$ for $\beta_{j}, \ j=1, ..., K$. 
\end{proposition}
\indent \textbf{Proof}. See Appendix B.2.

\indent In order to identify the unknown sign $\epsilon$, an additional constraint needs to be introduced such as
$$\beta_{1}>0.$$
\noindent The unknown sign $\epsilon$ is a problem of global identification and not of local identification. Hence, when the asymptotic properties of the estimators are derived (see Section 4), this positivity constraint has to be taken into account to obtain the consistency of the estimator. It has no effect on the asymptotic normality. The asymptotic properties of the composite log-likelihood estimators are discussed in the next section.

\section{Asymptotic Properties of Composite Log-likelihood Estimators}
\subsection{The Asymptotics}
\indent In a panel data framework, the asymptotic analysis can be performed with respect to the cross-sectional dimension $n$ and time dimension $T$ that can tend to infinity as follows.\footnote{The last case (iii) $ n$ fixed, $T \rightarrow{\infty}$ is less relevant for applications to credit rating.}
\begin{align*}
&(i) \  \mbox{Both }n,T\rightarrow{\infty}\mbox{: double asymptotics;}\\
&(ii) \ n\rightarrow{\infty}, \ T  \mbox{ fixed: short \ panel \ asymptotics.}
\end{align*}
The double asymptotics in case (i) has been developed for applications to big data [\citet{GG2015, GG2014}, \citet{BJR2017}]. It corresponds to a long panel of high dimensional time series.

When $n \rightarrow \infty, T \rightarrow \infty$, the asymptotic properties of the conditional maximum composite likelihood estimators are much easier to derive
than the asymptotic properties of the complete ML estimator. Indeed, the conditional composite log-likelihood functions are finite sums of products of summary statistics and functions of parameters. This simplifies the proof of uniform convergence with respect to the parameters. The next section examines the asymptotics (i) -(ii) and describes the properties  of the conditional composite maximum likelihood estimators.

\subsection{Consistency}

\indent This section examines the consistency of the maximum conditional composite likelihood estimators of the identifiable parameters when $n \rightarrow \infty, T \rightarrow \infty$. 
To prove the consistency, we need the following additional assumption.

\noindent {\bf Assumption A3}

a) The parameter set of $(\theta, \rho)$ is compact, and strictly included in the set
$\sigma_j>0, \forall j, |{\rho}|<1$.

b) The model is well-specified and the true value $(\theta_0, \rho_0)$ is in the interior of the parameter set.

The condition $\sigma_j>0, \; \forall j$, ensures that the transition probabilities
$p_{jk}(f_t; \theta)$ [resp. $p_{jk} (\theta), p_{jk}^{(2)} (\theta, \rho)$] are infinitely continuously
differentiable with respect to $f_t$ and $\theta$ (resp. with respect to $\theta, \rho$).

\noindent (i) Double asymptotics:  $n\rightarrow{\infty}, T\rightarrow{\infty} $

\indent Let us consider the double asymptotics with CL(1) approach. We have:

$$L_{cc}(\theta)=\sum_{k=1}^{K}\sum_{j=1}^{K}\sum_{t=2}^{T} \left[\pi_j \hat{p}_{jk,t} \ log \ p_{jk}(\theta)\right].$$

The conditional composite likelihood  $L_{cc}$ depends on $n$ and $T$, although it is not indexed by $n$ and $T$ to simplify the notation. Since $T$ is varying, we need uniform a.s. convergence of the ratios $\hat{p}_{jk,t}$ to $p_{jk}(f_t; \theta_0)$ with respect to $t$, not only their pointwise a.s. convergence.

\noindent {\bf Assumption A.4}

i) $P [Max_{m \geq n} | \hat{p}_{jk,t}(m) - p_{jk}(f_t; \theta_0)| > \epsilon \,|f_t) < \frac{g_{jk} (f_t; \theta_0)}{ n \epsilon^2}$, 

\noindent $\forall j,k,\epsilon, n, f_t$, where the notation $\hat{p}_{jk,t}(n)$ is introduced to indicate the dependence of the transition probabilities on the number of observations, and 
it exists a function $g_{jk}$ of $f_t$ and $\theta_0$, which
is integrable with respect to the marginal distribution of $f_t$.

ii) $n,T \rightarrow \infty$ with $T/n \rightarrow 0$.

\noindent Assumption A.4 i) is a domination condition. Assumption A.4 ii) means that we have  
a  panel with the cross-sectional dimension much larger than the time dimension. This allows for disregarding the uncertainty in $n$ with respect to the uncertainty in $T$ in the double asymptotic [see Appendix C].

\noindent Let us consider the objective function normalized by $T$,
$$\frac{1}{T} L_{cc}(\theta) = \sum_{j=1}^K  \pi_j \sum_{k=1}^K \left[ \left( \frac{1}{T} \sum_{t=2}^T 
\hat{p}_{jk,t}\right) \log \, p_{jk} (\theta) \right].$$
\noindent When $n$ tends  to infinity, this quantity tends to \\ 
\noindent $\sum_{j=1}^K  \pi_j \sum_{k=1}^K \left[ ( \frac{1}{T} \sum_{t=2}^T 
p_{jk}(f_t, \theta_0)) \log \, p_{jk} (\theta) \right].$
If moreover $T$ tends to infinity, the limiting objective function is

\begin{equation}
\lim_{n,T \rightarrow \infty} \frac{1}{T} L_{cc} \approx \sum_{j=1}^K  \pi_j \sum_{k=1}^K \left[ p_{jk}(\theta_0) \log \, p_{jk} (\theta) \right],
\end{equation}

\noindent by using the ergodicity of the factor process and the Strong Law of Large Numbers in time dimension.

\indent By the property of the Kullback-Leibler divergence measure applied to each row of the transition matrix, we know that the associated limiting conditional composite log-likelihood is maximized at $\theta_0^*$, with
$$p_{jk}(\theta_0^*)=p_{jk}(\theta_{0}), \forall j,k.$$ 
\noindent Then, by the identifiability of $\theta=(c, \delta, \gamma)$ (see Proposition 1), we get $\theta^{*}_{0}=\theta_{0}$, and the consistency follows.

\noindent (ii) Short panel asymptotics: $n\rightarrow\infty,T$ fixed.

\indent Like for the granularity approach, we cannot expect the conditional composite ML estimators to be consistent for $n \rightarrow \infty$, $T$ fixed. This is a consequence of the cross-sectional dependence due to the common systemic factor $f_{t}$. To clarify this point, let us assume $T=2$ and consider the maximum conditional composite likelihood  CL(1) estimator. For $T=2$, the conditional composite log-likelihood is
$$L_{cc}(c,\delta,\gamma)=\sum_{k=1}^{K}\sum_{j=1}^{K} \left[\pi_j \hat{p}_{jk,2} \ log \ p_{jk}(\theta)\right],$$
where $\theta = (c, \delta, \gamma)$ is the identifiable parameter satisfying the identification restriction in Proposition 1, that are $c_2=0,\gamma_1=1$. By Assumptions A.1, A.2 and the fact that the rating indicators are nonnegative and bounded, we can apply the Strong Law of Large Numbers to individuals. The conditional composite log-likelihood tends a.s. to
$$\lim_{n\rightarrow{\infty}}  \mbox{a.s.}   L_{cc}(c,\delta,\gamma)=\lim_{n\rightarrow{\infty}} \mbox{a.s.} \sum_{j=1}^{K} \pi_j \Bigg( \sum_{k=1}^{K} \Bigg[ p_{jk} (\theta_0, f_{02}) \ log \ p_{jk}(\theta) \Bigg] \Bigg),$$

Then, this limiting objective function admits at least a maximum on the parameter set by Assumption A4 ii). Let  $ \theta_0^*$ denote the pseudo-true value, i.e. a solution of the asymptotic optimization problem, we have
$$ \theta_0^* = \underset{\theta}{\operatorname{argmax}}\sum_{j=1}^{K} \Bigg[ \pi_j \bigg[ \sum_{k=1}^{K} p_{jk}(\theta_0,f_{0,2}) \ log \ p_{jk}(\theta) \bigg] \Bigg]. $$
This pseudo-true value is a function of $\theta_{0}$ and $f_{0,2}$. Therefore, it cannot be equal to the true value, that does not depend on $f_{0,2}$. In other words, the MCL estimator $\hat{\theta}_n$ converges to a stochastic limit whose distribution depends on the distribution of $f_2$.

\subsection{Asymptotic Normality}
For expository purpose, we continue the discussion of the CL(1) approach for $n \rightarrow \infty, T \rightarrow \infty$. 
As mentioned above, the conditional composite log-likelihood is continuously differentiable. Since the estimator $\hat{\theta}_{n,T} = (\hat{c}_{n,T}, \hat{\delta}_{n,T}, \hat{\gamma}_{n,T})$ tends to the true value $\theta_0= (c_0, \delta_0, \gamma_0)$, which is in the interior of the parameter set, the estimator will also be asymptotically in the interior of the parameter set and will satisfy the necessary first-order conditions for large $T$. Therefore, we have
$$
\frac{\partial L_{cc} (\hat{\theta}_{n,T})}{\partial \theta}=0
\iff \sum_{k=1}^{K}\sum_{j=1}^{K}\sum_{t=2}^{T} \left[ \pi_j \hat{p}_{jk,t} \ \frac{\partial \ log p_{jk}(\hat{\theta}_{n,T})}{\partial \theta}\right]=0.
$$
We can perform a Taylor-McLaurin expansion with respect to $\hat{\theta}_{n,T}$  in the neighborhood of $\theta_{0}$.
Let us assume: 

\noindent {\bf Assumption A.5} 
The parameter set $\Theta$ for $\theta$ is convex.

 We get 
\begin{equation}
 \sum_{k=1}^{K}\sum_{j=1}^{K}\sum_{t=2}^{T} \left[\pi_j \hat{p}_{jk,t} \ \frac{\partial log \ p_{jk} (\theta_{0})}{\partial \theta}\right]+\left(\sum_{k=1}^{K}\sum_{j=1}^{K}\sum_{t=2}^{T} \left[ \pi_j \hat{p}_{jk,t} \ \frac{\partial^{2} log \ p_{jk} (\tilde{\theta}_{n,T})}{\partial \theta \ \partial \theta^{'}} \right] (\hat{\theta}_{n,T}-\theta_{0}) \right)= 0,
\end{equation}
\noindent where $\tilde{\theta}_{n,T}$ is an intermediate value between $\hat{\theta}_{n,T}$
and $\theta_0$. 

By applying the same argument as for the uniform a.s. convergence of the composite log-likelihood function, we deduce that

\noindent $\frac{1}{T} \sum_{k=1}^K \sum_{j=1}^K \sum_{t=2}^T \left[\pi_j \hat{p}_{jk,t} \frac{\partial^2 log  \ p_{jk} (\tilde{\theta}_{n,T}) }{ \partial \theta \partial \theta'} \right]$ will converge a.s. to 
$\sum_{k=1}^K \sum_{j=1}^K \left[\pi_j p_{jk} (\theta_0)  \frac{\partial^2 log  \ p_{jk} (\theta_0) }{ \partial \theta \partial \theta'}\right]$, 

\noindent $\frac{1}{T} \sum_{k=1}^K \sum_{j=1}^K \sum_{t=2}^T \left[\pi_j \hat{p}_{jk,t} \frac{\partial log p_{jk} (\theta_0) }{ \partial \theta}\right]$ will converge a.s. to

$ \sum_{k=1}^K \sum_{j=1}^K \left[\pi_j p_{jk} (\theta_0) \frac{\partial log  \ p_{jk} (\theta_0) }{ \partial \theta}\right]=0$,

\noindent since $\theta_0$ is the maximizer of the limiting objective function (4.1), and
\begin{eqnarray*}
\lefteqn{ \frac{1}{ \sqrt{T}} \sum_{k=1}^K \sum_{j=1}^K \sum_{t=2}^T \left[\pi_j \hat{p}_{jk,t} \frac{\partial log \ p_{lk} (\theta_0) }{ \partial \theta}\right]} \\
& = & \frac{1}{ \sqrt{T}} \sum_{k=1}^K \sum_{j=1}^K  \left\{ \left[ \sum_{t=2}^T \pi_j  [p_{jk}(f_t, \theta_0) - p_{jk}(\theta_0)] \right] \frac{\partial log  \ p_{jk} (\theta_0) }{ \partial \theta} \right\} +o_p(1),
\end{eqnarray*}
where $o_p(1)$ is a negligible term in probability by Assumption A.4.
Let us assume:

\noindent {\bf Assumption A.6} 
The matrix $J_0 = \sum_{k=1}^K \sum_{j=1}^K \left[ \pi_j  p_{jk} (\theta_0) \frac{\partial^2 log \  p_{jk} (\theta_0) }{ \partial \theta \partial \theta'} \right]$ is positive definite.

Then, by normalizing the expansion (4.2) by $1/( \sqrt{T})$, we get
\begin{eqnarray*}
& \sqrt{T} (\hat{\theta}_{n,T}-\theta_{0}) = \Bigg [-\sum_{k=1}^{K}\sum_{j=1}^{K}\left( \pi_j  \ p_{jk}(\theta_{0}) \ \frac{\partial^{2} log \ p_{jk} (\theta_{0})}{\partial \theta \partial \theta' }\right)\Bigg]^{-1}  \\
&\times\frac{1}{\sqrt{T}} \sum_{t=2}^{T} \sum_{k=1}^K \sum_{j=1}^K \left[ \pi_j [p_{jk} (f_t; \theta_0)
- p_{jk}(\theta_0)] 
\frac{\partial log \ p_{jk}(\theta)}{\partial \theta} \right] + o_p(1)  \\
& = \left[ - \sum_{k=1}^K \sum_{j=1}^K  \pi_j p_{jk}(\theta_0) \frac{\partial^2 log \ p_{Jk}(\theta_{0})}{\partial \theta \partial \theta'} \right]^{-1} \frac{\partial}{\partial \theta}
\left[ vec \log p_{jk}(\theta_0)\right]' \\
& \times \frac{1}{\sqrt{T}} \sum_{t=2}^T vec [\pi_j [p_{jk}(f_t, \theta_0)- p_{jk}(\theta_0)]] + o_p(1),
\end{eqnarray*}
where $vec$ denotes the vectorization that stacks the columns of the transition matrix.
Note that 
$$ vec [\pi_j p_{jk} (f_t; \theta_0)] =  vec[P(f_t; \theta_0) diag \, \pi] = [diag \, \pi \otimes Id] \, vec P(f_t; \theta_0),$$

\noindent where $diag \, \pi$ is the diagonal matrix with terms $\pi_j$ on the main diagonal and $\otimes$ denotes the Kronecker product.

The common factor $f_{t}$ is strictly stationary and geometrically mixing. Thus, the same property holds for the $K^{2}$ dimensional process $ vec\left[\pi_j  p_{jk} (f_t; \theta_0)\right]$. We deduce the asymptotic normality of 
\begin{align*}
    \sqrt{T}(\hat{\theta}_{n,T}-\theta_{0}).
\end{align*}

\begin{proposition}
Under Assumptions A.1-A.6, when $n \rightarrow \infty, T \rightarrow \infty$, the maximum conditional composite likelihood estimator $\hat{\theta}_{n,T}$ obtained by maximizing $L_{cc}(\theta)$, defined in equation (3.3), is consistent, converges to the true value $\theta_{0}$ at speed $1/\sqrt{T}$, and is asymptotically normal,
\begin{align*}
 \sqrt{T}\left(\hat{\theta}_{n,T}-\theta_{0}\right) \sim N \bigg[ 0, J^{-1}_{0}\left(\sum_{h=-\infty}^{\infty} I_{0h}\right) J_{0}^{-1}\bigg],
\end{align*}
\noindent where
\begin{align*}
    &J_{0}= - \sum_{k=1}^{K}\sum_{j=1}^{K}\left[\pi_j p_{jk}(\theta_{0}) \ \frac{\partial^{2} log \ p_{jk} (\theta_{0})}{\partial \theta \partial \theta^{'} }\right],\\
    &I_{0h}= \frac{\partial}{\partial \theta} vec\left(\log p_{jk} (\theta_0) \right)^{'} Cov_{0}\bigg[vec\left(\pi_j p_{jk} (f_t, \theta_0) \right), vec(\pi_j p_{jk} (f_{t-h},\theta_0)) \bigg] \\
    &\  \  \  \  \  \  \times \frac{\partial}{\partial \theta'} vec\left(\log p_{jk} (\theta_0) \right),\\
    &\ \ \ \ =\frac{\partial}{\partial \theta} vec[ \log p_{jk}(\theta_0)]' ( diag \, \pi \otimes Id) Cov_0 [vec P (f_t, \theta_0), vec P (f_{t-h}, \theta_0)] \\
    & \ \ \ \ \ \times (diag \, \pi \otimes Id) \frac{\partial}{\partial \theta'} vec[ \log p_{jk}(\theta_0)], \\
    & \ \ \ \ h=1, 2, ...
\end{align*}
\end{proposition}
As expected, we obtained the following results.

\noindent (a) \ The speed of convergence of $\hat{\theta}_{n,T}$ is $1/\sqrt{T}$ instead of $1/\sqrt{nT}$, which characterizes the granularity approach. This is a consequence of the crude cross-sectional aggregation of the data in the composite approach as if the observations $y_{i,t}$ were cross-sectionally independent.
However, a drawback of the granularity approach is the need to estimate the $T-1$ latent factors, which increases its computational complexity compared to the conditional MCL methods.

\noindent (b) \ The asymptotic variance is obtained from the ``sandwich'' formula, as it is common in a mis-specified (pseudo) maximum likelihood approach [see, \citet{Godambe1960}, \citet{Hubert1967}, \citet{White1982}].

\noindent (c) \  The terms $p_{jk} (f_t, \theta_0)$ and $p_{jk}(f_{t-h}, \theta_0)$ depend on $f_{t}$, and $f_{t-h}$, respectively. They are correlated because of the factor dynamics (except when $\rho=0$, that is the case of an i.i.d. factor). Therefore, the covariances have to be taken into account even if we consider only a small number of values of lag $h$. It is important to notice that the sum $\sum_{h=-\infty}^{\infty} I_{0h}$ 
always exists due to the geometric ergodicity of the factor process.

\noindent (d) \ The asymptotic variance-covariance matrix of the MCL estimator depends on the selected set of weights $\pi$. It is out of the scope of this paper to discuss the optimal choice of weights that likely reduce the robustness of this estimator. Instead, to facilitate the comparison with the granularity-based approach, this set of weights has to be close to the structure of ratings at the different dates. As it is assumed to be time independent, a solution is to choose the set of weights close to the true unconditional structure of ratings. 
More specifically, we choose the fixed rating structure to be the average of the observed rating structures across time in the simulation and the empirical application. For the different states, these observed rating structures are given by the number of firms in each state at the end of each time period divided by the total number of firms transiting from non-default states. The simulation results in Appendix D suggest that this choice yields rating structures that can approximate the unknown unconditional rating structure (see Sections 5 and 6).

\indent The above asymptotic analysis is different from the main literature on composite likelihood that usually considers either i.i.d. individuals, or finite dimensional time series [see e.g. \citet{CR2004}, \citet{VRF2011}]. 

\indent The asymptotic variance-covariance matrix of the conditional composite maximum likelihood estimator is consistently estimated by considering appropriate sample counterparts of components $J_0, I_{0h}$. 

\section{Simulation Results}

In this section, we perform a Monte-Carlo experiment to assess the finite sample properties of estimators based on the conditional composite likelihood function and step one of the granularity approach.

\subsection{The Design}

The designs include $K=8$ ratings, with a higher $k$ indicating a lower capacity to repay debt, and $k=8$
denoting the absorbing state of default. These rating categories can be interpreted as AAA, AA, A, BBB, BB, B, CCC/CC and D, respectively (according to the Standard and Poor's (S \& P) terminology).

\subsubsection{Design of thresholds and intercepts}

Given the rating at time $t-1,$ i.e. $y_{i,t-1}=j\in\left\{  1,\ldots
,7\right\}  $,  suppose that the underlying latent continuous quantitative score
$y_{i,t}^{\ast}$ can be written as
\[
y_{i,t}^{\ast}=\delta_{j}+\beta_{j}f_{t}+\sigma_{j}u_{i,t},\text{ }u_{i,t}\sim
i.i.d.N\left(  0,1\right)  ,
\]
where the rating is determined by
\[
y_{i,t}=k,k=1,\ldots,8\Longleftrightarrow c_{k}\leq y_{i,t}^{\ast}%
<c_{k+1},k=1,\ldots,8,
\]
with the thresholds $\left(  c_{k}\right)  $ described in Table 1 and the
intercepts $\left(  \delta_{j}\right)  $ described in Table 2.

{\renewcommand{\arraystretch}{1.0}
} \begin{table}[ptbh]
\caption{Thresholds $\left(  c_{k}\right)  $}
\label{Tab1}
%
\centering
\begin{tabular}
[c]{c|c|c|c|c|c|c|c|c|c}\hline\hline
$k$ & $1$ & $2$ & $3$ & $4$ & $5$ & $6$ & $7$ & $8$ & $9$\\\hline\hline
$c_{k}$ & $-\infty$ & $0$ & $1.5$ & $3$ & $4.5$ & $6$ & $7.5$ & $9$ & $\infty
$\\\hline\hline
\end{tabular}
\end{table}

{\renewcommand{\arraystretch}{1.0}
} \begin{table}[ptbh]
\caption{Intercepts $\left(  \delta_{j}\right)  $}%
\label{Tab2}
%
\centering
\begin{tabular}
[c]{c|c|c|c|c|c|c|c}\hline\hline
$j$ & $1$ & $2$ & $3$ & $4$ & $5$ & $6$ & $7$\\\hline\hline
$\delta_{j}$ & $-0.5$ & $1$ & $2.5$ & $4$ & $5.5$ & $7$ & $8.5$\\\hline\hline
\end{tabular}
\end{table}

\noindent The thresholds and intercepts are ranked in an increasing order, and their values are chosen to get higher transition probabilities on the
main diagonal and decreasing probabilities when a firm transits to other
states. The treatment of the ``absorbing state D'' corresponding to $j=8$ is discussed later in Section 5.1.3.

\subsubsection{Design of risk components}

The uncertainty on migrations is driven by rating-specific shocks $u_{i,t}$ and
the common systematic shocks $f_{t}$. To see the effects of
risk on the systematic and  rating-specific components, we consider two
designs for $\sigma_{j},\beta_{j}, j=1,\ldots,7$.

 \noindent Design 1: $\rho=0$; The idiosyncratic and systemic components have, for each $j$, the same impact, that is: $\sigma
_{j}=\beta_{j}=\frac{(1+r)^{j-1}}{\sqrt{2}}$, with $r=0.05$. Thus, when the rating is lower, the risk of downgrading is higher.

  \noindent  Design 2: $\rho = 0.4, 0.7$ and $0.95$; The autocorrelation parameter is taken into account and the impact of the systemic component relative to the
idiosyncratic one decreases with $l$. This means that the idiosyncratic errors largely explain the junk bonds in non crisis environment. To capture this feature, we consider the
ratios $\frac{\beta_{j}}{\sigma_{j}}=\frac{1}{(1+r)^{j-1}}$, with $r=0.05$, where $\beta
_{j}=\frac{1}{\sqrt{2-\rho^{2}}}$, $\forall j$.

  \noindent There is also a persistence of the systematic factor $f_{t}$ satisfying
\[
f_{t}=\rho f_{t-1}+\sqrt{1-\rho^{2}}\eta_{t}, \eta_{t}\sim i.i.d.N(0,1),
\]
where the autocorrelation parameter $\rho$ measures the persistence and $f_{1}$ is drawn in th estationary distribution $N(0,1)$.

\noindent We consider the following four values for the autocorrelation parameter $\rho$: 
$\rho=0$, that corresponds to independent migration matrices. This is the
basic assumption of the Value of the Firm model introduced in \citet{Vasicek1987}; 
$\rho=0.4$ is used to reflect a moderate amount of autocorrelation at lag 1 of the systematic factor; $\rho=0.7$ corresponds to a high amount of autocorrelation at lag 1 of the systematic factor, while $\rho=0.95$ allows for some persistence in the systematic factor.

\subsubsection{Treatment of the absorbing state}

The state of default is an absorbing state. Therefore, if
we follow a given population of corporates, all of the corporates
will default at some date, and the number of still alive corporates
(the so-called Population-at-Risk (PaR)) will diminish.
Theoretically, the process of observed ratings is asymptotically
stationary with a stationary distribution equal to a point mass on default.
This difficulty is solved by assuming
that newly created corporates
offset the corporates entering into default, thus ensuring a PaR of
constant size. This corresponds to the model with equal birth
and death rates used in epidemiological studies (see e.g. \citet{HLM2014}).
As at the time of new firms arrival their rating are high, we replace the last row of the
migration matrix at the individual level,

\[
0, 0, 0,0,0,0,0,1,
\]
corresponding to a standard absorbing state, by the row of assignment of new entries at the population level,
\[
0.5,0.3,0.2,0,0,0,0,0.
\]

Thus we have to distinguish individual migration matrices $P_{t}$,
from the population migration that could be adjusted by taking into
account the newly created firms. When the newly created firms are taken into account,
the migration matrix is indexed as $P^{a}_t$.

\subsubsection{Individual trajectories}

Let us consider the design with $\rho=0.4$. For each individual $i$, we compute and
compare the time series of underlying scores,  ratings as well as the series of expected stability measures
in the
current rating. These series are denoted by $y_{i,t}^{\ast}$, $y_{i,t}$ and $s_{i,t}$,
where
\[
s_{i,t}=\Phi\left(  \frac{c_{k+1}-\beta_{j}f_{t}+\delta}{\sigma_{j}}\right)
-\Phi\left(  \frac{c_{k}-\beta_{j}f_{t}+\delta}{\sigma_{j}}\right)  ,\text{
with }y_{i,t-1}=j.
\]
These series are displayed in Figure 1 for an initial factor
value of $f_{1}=0$ and initial rating of $y_{i,0}=2$, which is equivalent to AA.
\begin{figure}[ptbh]
\caption{Individual Trajectories, $\rho=0.4$}%
\label{fig2}%
\includegraphics[width=\linewidth]{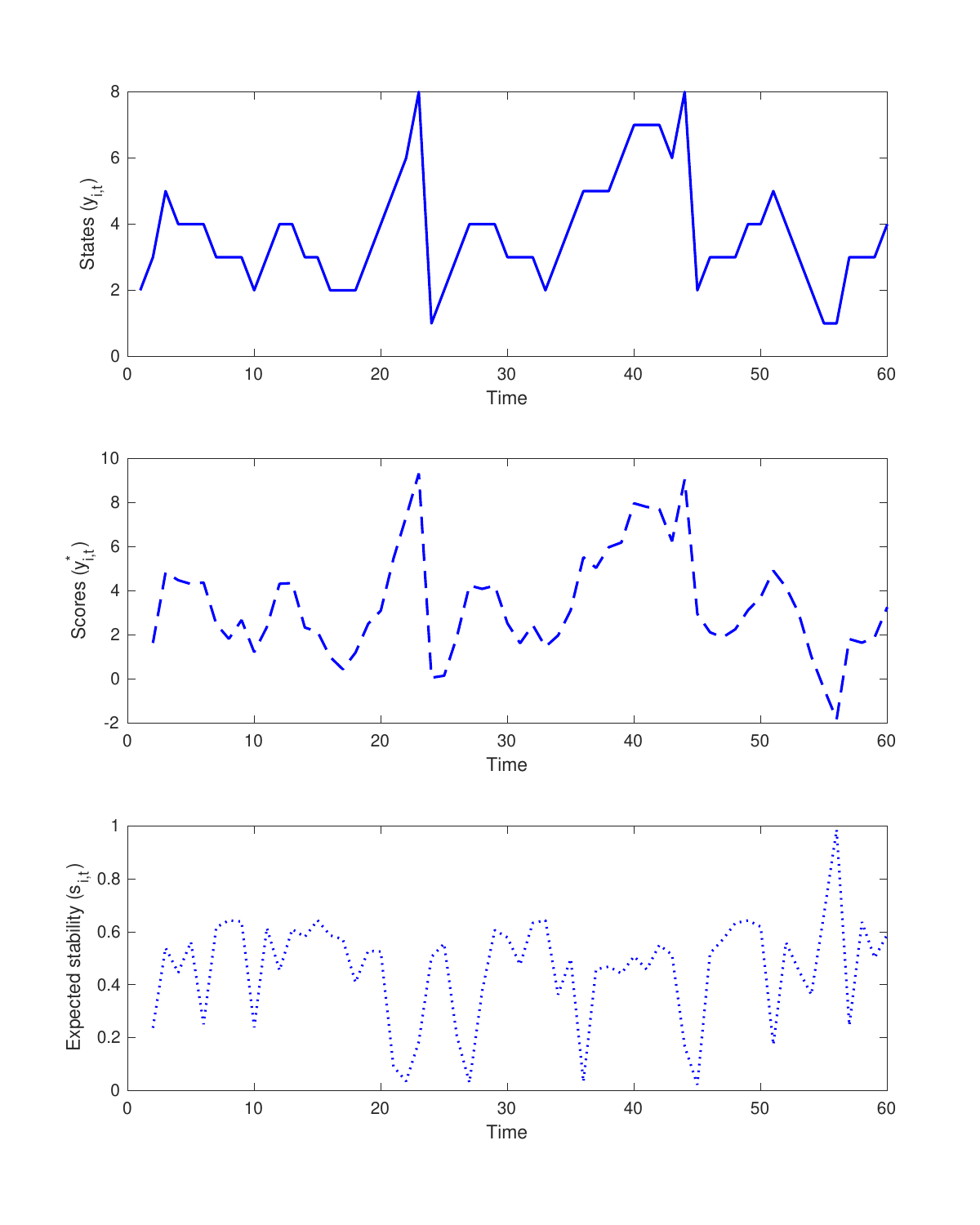}
\end{figure}
\bigskip

The displayed trajectories correspond to three different corporate bonds. At time 0, a bond with rating 2 (AA) is issued.
It is subject to downgrading after time 10 down to default in time 21. At that time a new bond with rating 1 (AAA) is issued to balance the defaulted bond. It is gradually downgraded to default at time 44.  Then, a new bond is issued at time 45 and so on.
In such an environment of births and deaths occurring with equal rates, each trajectory corresponds to a stochastic number of firms, rather than a single firm. This stochastic number is equal to the number of observed defaults plus one. This approach ensures the stationarity of the process and provides the rating histories of equal length $T$. 

\subsubsection{Quasi-Migration Matrices}

In this section we present the quasi-transition (migration) matrices with $\rho=0.4$. The time unit can be viewed as one month or one quarter,  and the horizons of one and two correspond to one and two time units, respectively.  Matrix $P^{a}$ is evaluated at the true parameter value from the formula in Lemma 1 and given in Table 3.

{\renewcommand{\arraystretch}{0.9}
\begin{table}[ptbh]
\caption{ Quasi-Migration Matrices $P^{a}$, at Horizon 1 in \%}%
\label{Tab3}
\bigskip
\centering%
\begin{tabular}
[c]{c|c|c|c|c|c|c|c|c}\hline\hline
$P^{a}$& $k=1$ & $k=2$ & $k=3$ & $k=4$ & $k=5$ & $k=6$ & $k=7$ & $k=8$\\\hline\hline
$j=1$ & $68.42$ & $28.82$ & $2.72$ & $0.04$ & $0.00$ &$0.00$ & $0.00$ & $0.00$\\
$j=2$ & $17.48$ & $50.53$ & $28.93$ & $3.01$ & $0.05$ & $0.00$ & $0.00$ & $0.00$\\
$j=3$ & $1.14$ & $16.97$ & $49.46$ & $29.01$ & $3.35$ &$0.07$ & $0.00$ & $0.00$\\
$j=4$ & $0.02$ & $1.31$ & $17.43$ & $48.36$ & $29.07$ & $3.71$& $0.10$ & $0.00$\\
$j=5$ & $0.00 $ & $0.03$ & $1.53$ & $17.88$ & $47.23$ &$29.09$ & $4.11$ & $0.13$\\
$j=6$ & $0.00$ & $0.00$ & $0.04$ & $1.78$ & $18.32$ &$46.07$ & $29.07$ & $4.72$\\
$j=7$ & $0.00$ & $0.00$ & $0.00$ & $0.06$ & $2.07$ & $18.73$ & $44.89$ & $34.25$\\
$j=8$ & $50.00$ & $30.00$ & $20.00$ & $0.00$ & $0.00$ & $0.00$ & $0.00$ & $0.00$\\\hline\hline
\end{tabular}
\end{table}}

We observe a commonly reported feature of a migration matrix, i.e. the largest rates located on the main diagonal and the two adjacent diagonals with larger rates of downgrades than of upgrades. 
Moreover, there are significant rates of default from ratings 6 and 7, corresponding to the ``junk bonds''. The last row corresponds to the new firms introduced to compensate for the defaulted corporates. Next, we determine the nondegenerate stationary distribution $\mu^{a}$, solution of
\begin{equation}
\left(  \mu^{a}\right)  ^{\prime}=\left(  \mu^{a}\right)  ^{\prime}P^{a}.
\end{equation}

Because of the absorbing state, without the equal birth-death rates each corporate bond would default and the asymptotic stationary distribution of individual ratings would be a point mass at 8 (D). The interpretation of the stationary distribution $\mu^a$ is different and concerns the population ratings. It provides the long run rating structure of the population of corporate bonds under rebalancing. This long run structure, that does not depend on the initial rating structure, is given in Table 4. In practice, the stationary distribution provides the information on how the ratings agencies determine the thresholds of scores to define the ratings. In our experimental design, the unobserved quantitative scores are discretized to obtain close proportions of bonds across ratings.
{\renewcommand{\arraystretch}{0.9}
\begin{table}[ptbh]
\caption{ Stationary Distribution}%
\label{Tab4}%
\bigskip
\centering%
\begin{tabular}
[c]{c|c|c|c|c|c|c|c|c}\hline\hline
& $k=1$ & $k=2$ & $k=3$ & $k=4$ & $k=5$ & $k=6$ & $k=7$ & $k=8$\\\hline\hline
Probabilities in \% & $14.51$ & $16.66$ & $17.47$ & $16.09$ & $14.15$ &
$11.19$ & $6.99$ & $2.94$\\\hline\hline
\end{tabular}
\end{table}}

\noindent Let us now consider the quasi-migration matrix at horizon
2. Table 5 shows the matrices $P^{(2)a}$ and $(P^a)^{2}$. Matrix $P^{(2)a}$ is computed by Monte-Carlo integration with $S=50,000$ replications of the latent factor values $f_t$ at the true parameter value (see Lemma 2).
The changes in $P^{(2)a}$ and $(P^{a})^{2}$ compared to $P^a$ are due to
the aggregate effect of both the rating-specific and systemic
shocks. Both matrices have non zero elements on the
diagonals up or down by 2 from the main diagonal because of time
aggregation. 
The matrices $P^{(2)a}$
and $(P^{a})^{2}$ are not equal. This difference 
is caused by the systemic risk. As expected, we observe larger diagonal elements in the matrix $P^{(2)a}$ for $j=2,...,7$ because of the persistence of the factor, which leads to more stability in the ratings.

{\renewcommand{\arraystretch}{0.9}
\begin{table}[ptbh]
\caption{Quasi-Migration Matrices, at Horizon 2 in \%}%
\label{Tab5}%
\centering%
%
\begin{tabular}
[c]{c|c|c|c|c|c|c|c|c}\hline\hline
$P^{a}(2)$& $k=1$ & $k=2$ & $k=3$ & $k=4$ & $k=5$ & $k=6$ & $k=7$ & $k=8$\\\hline\hline
$j=1$ & 51.89 & 34.75 & 11.54 & 1.70 & 0.12 & 0.00 & 0.00 & 0.00\\
$j=2$ &   21.12 & 35.51 & 29.93 & 11.39 & 1.90 & 0.15 & 0.00 & 0.00\\
$j=3$ & 4.31 & 17.68 & 34.51 & 29.49 & 11.69 & 2.12 & 0.19 & 0.01\\
$j=4$ &  0.45 & 4.27 & 17.88 & 33.75 & 29.05 & 11.99 & 2.36 & 0.25\\
$j=5$ & 0.09 & 0.56 & 4.64 & 18.06 & 32.97 & 28.58 & 12.26 & 2.84\\
$j=6$ & 2.36 & 1.45 & 1.57 & 4.99 & 18.21 & 32.07 & 27.20 & 12.15\\
$j=7$ & 17.13 & 10.28 & 6.90 & 0.76 & 5.35 & 17.64 & 25.68 & 16.26\\
$j=8$ & 39.68 & 32.96 & 19.93 & 6.73 & 0.69 & 0.01 & 0.00 & 0.00 \\\hline
\end{tabular}
\begin{tabular}
[c]{c|c|c|c|c|c|c|c|c}\hline\hline
$(P^{a})^2$& $k=1$ & $k=2$ & $k=3$ & $k=4$ & $k=5$ & $k=6$ & $k=7$ & $k=8$\\\hline\hline
$j=1$ & 52.90 & 31.85 & 12.59 & 2.40 & 0.25 & 0.01 & 0.00 & 0.00\\
$j=2$ & 22.83 & 33.32 & 28.37 & 12.56 & 2.61 & 0.29 & 0.02 & 0.00\\
$j=3$ &  5.61 & 17.88 & 32.51 & 28.06 & 12.74 & 2.83 & 0.35 & 0.02\\
$j=4$ &  0.76 & 5.23 & 18.03 & 31.82 & 27.72 & 12.92 & 3.08 & 0.44\\
$j=5$ &  0.13 & 0.86 & 5.56 & 18.16 & 31.13 & 27.33 & 13.09 & 3.74\\
$j=6$ & 2.36 & 1.49 & 1.89 & 5.85 & 18.26 & 30.38 & 26.33 & 13.44\\
$j=7$ & 17.18 & 10.31 & 6.97 & 1.10 & 6.17 & 17.84 & 24.94 & 15.49\\
$j=8$ & 39.64 & 32.98 & 19.94 & 6.74 & 0.69 & 0.01 & 0.00 & 0.00\\\hline\hline
\end{tabular}
\end{table}}

\subsection{Finite Sample Properties of the MCL Estimation}
To give some insights into the accuracy of the MCL estimation, in
terms of the number of months $T$ and the factor autocorrelation
parameter $\rho$, we conduct  Monte-Carlo experiments.
The estimation is performed with $n=1,000$ firms, including the adjustment for
the newly created firms and the designs described above. The numbers
of observation periods are $T=60$ ($5$ years for monthly data and $15$ years for quarterly data), $T=120$ ($10$ years for monthly data and $30$ years for quarterly data),
$T=240$ ($20$ years for monthly data and $60$ years for quarterly data). In each experiment, we perform $S=500$ simulations of
individual trajectories, with initial ratings $y_{i,0}$ drawn from the
adjusted stationary distribution $\mu^{a}$, 
conditional on non-default ratings. The stationary probability computed conditional on non-default ratings provides the initial rating structure.
The condition for this initial draw is satisfied by starting the simulations from another initial structure, $20$ dates before $t=1$ in our case. This structure is chosen equal to the average observed structure of ratings.\footnote{Figures  \ref{Ratings_T60_rho_1} - \ref{Ratings_T240_rho_4} in online Appendix D.4 presents the rating structures at each time period $t$, and the chosen fixed rating structure, represented by their means over the $S$ simulated data. The figures show that the rating structures at each date $t$ are generally close to the fixed structure for most ratings. We find an improvement in the ability of the average rating structures to capture their dynamics over time as the autocorrelation in the systematic factor increases. 
However, as mentioned above, the choice of optimal rating structure is beyond the scope of this paper.}

\subsubsection{Parameters of Interest}
The stochastic migration model depends on a large number of parameters that are $19$ identifiable parameters  ($c_{3}\ldots,c_{8}$, $\delta_{1}\ldots,\delta_{7}$, $\gamma_{2}\ldots\gamma_{7}$) for the CL(1) method, where $\gamma_j=\sqrt{\sigma_j^2+\beta_j^2}$, and $27$ identifiable parameters ($c_{3}\ldots,c_{8}$, $\delta_{1}\ldots,\delta_{7}$,  $\beta_{2}\ldots,\beta_{7}$, $\sigma_{1}\ldots,\sigma_{7}$ and $\rho$) for the CL(2) method, counted by taking into account the two identification conditions $c_{2}=0,\gamma_{1}=1$. For the granularity approach, further identification conditions are needed.  The next sections discuss the  CL(1) estimation results based on design 1, the CL(2)  estimation results based based on design 2, and compare the findings to the granularity estimation results. In particular, we discuss the finite sample distributions of the estimated $c_{k+1},k=2,\ldots,7$ and  $\delta_{j},j=2,\ldots,7,$ which are common for the CL(1), the CL(2) and the granularity  methods. In addition,  we analyze the distribution of the estimated $\beta_{j},j=2,\ldots,7$ and  $\sigma_{j},j=2,\ldots,7,$ obtained from the CL(2) and the granularity  approaches.

\subsubsection{The CL(1) Estimation Results}
The CL(1) approach depends on the selected set of weights $(\pi_j,\; j=1,...,7)$ with weight zero on the default rating. 
This set of weights is set equal to the average rating structure from 
the simulated data conditional on non-default ratings.
We illustrate the finite sample behavior of the parameter estimators by plotting the empirical pdf of the estimates (Figures \ref{CL1_thresholds1} - \ref{CL1_variances4} in online Appendix D.1), when $\rho=0, 0.4, 0.7$ and $0.95$ for $T=60, 120$ and $240$. Figures \ref{CL1_thresholds1} - \ref{CL1_thresholds4} present the empirical probability distributions depicted by the histograms for the estimated threshold parameters $\left(c_{3},\ldots,c_{8}\right)$. Figures \ref{CL1_intercepts1} - \ref{CL1_intercepts4} show the empirical probability distributions of the estimated intercepts $\left(  \delta_{1},\ldots,\delta_{7}\right)  $, while Figures  \ref{CL1_variances1} - \ref{CL1_variances4}  present these distributions for the unconditional variances $\left(\gamma_{2},\ldots,\gamma_{7}\right)  $. In each figure, the $x$-axis shows the values of estimators, while the $y$-axis presents their frequencies. The red vertical line shows the true value of the estimated parameters and allows us to analyze to which extent the estimators are biased and discuss their distributions. When $\rho=0, 0.4$ and $0.7,$ a common feature observed in these figures is that, when $T$ varies, the distribution of the parameter estimators remains centered around their true values. The mode of estimates from the simulated data takes values close to the true value.  As the sample sizes increase, the range of values taken by the estimates tends to decrease. This indicates a smaller dispersion, and, therefore, an improvement in the precision of the CL(1) estimation. 
In the extreme case, where $\rho=0.95$ is close to the non-stationarity in the latent factor, with $\rho=1$, the estimations are less accurate. However, the accuracy also improves with an increase in the time dimension. The results are in line with the asymptotic results on the $\sqrt{T}$-consistency of the CL(1) estimates given in Proposition 3. 

\subsubsection{The CL(2) Estimation Results}
We conduct further analysis based on the CL(2) method with the same set of weights.  Figures \ref{CL2_thresholds1} - \ref{CL2_variances4} in online Appendix D.2 show the distributions of the CL(2) estimators when $\rho=0, 0.4, 0.7$ and $0.95$ for $T=60, 120$ and $240$.  Like for the CL(1), Figures \ref{CL2_thresholds1} - \ref{CL2_thresholds4}
provide the histograms of the estimated thresholds, and Figures \ref{CL2_intercepts1} - \ref{CL2_intercepts4} show the histograms of the estimated intercepts. In addition to those estimated parameters, Figures \ref{CL2_slopes1} - \ref{CL2_slopes4} display the histograms of the estimated slopes, while Figures  \ref{CL2_variances1} - \ref{CL2_variances4}  present these distributions for the estimated volatilities.
Under the CL(2), the thresholds and intercepts are generally well estimated when $\rho=0, 0.4$ and $0.7$. The results show that the estimates are centered around the true parameters. The accuracy of the estimation improves with increasing sample sizes. We observe similar results for the factor sensitivities and  volatilities. 

\subsubsection{The Granularity Estimation Results} 
The granularity estimation consists of two steps. First,  the log-likelihood from the micro-density given by
\begin{equation}
\sum_{t=2}^T \sum_{j=1}^K \sum_{k=1}^K n_{jk,t} log \big[ \Phi\big(\frac{c_{k+1}-\beta_j f_t -\delta_j}{\sigma_j}\big)- \Phi\big(\frac{c_{k}-\beta_j f_t -\delta_j}{\sigma_j}\big)\big],
\label{step1}
 \end{equation}
is maximized with respect to parameter $\theta$ and factor values $f_2,...,f_T$. 
In the second step, the values $\hat{f}_t$ are regressed on their lagged values to get an estimator of $\rho$.  
We examine the first step of the granularity approach providing the estimates of $\theta$. 
 
The estimators are presented in online Appendix D.3. Figures \ref{GG_thresholds1} - \ref{GG_thresholds4} provide the histograms of the estimated thresholds, and Figures \ref{GG_intercepts1} - \ref{GG_intercepts4} show the histograms of the estimated intercepts. Figures \ref{GG_slopes1} - \ref{GG_slopes4} show the histograms of the estimated slopes, while Figures  \ref{GG_variances1} - \ref{GG_variances4}  present these distributions for the estimated volatilities. By comparing the finite sample distributions, we observe that
the CL(1) method provides slightly more accurate results than the CL(2) and the first step of the granularity approach for the parameters identifiable under the CL(1).
The results obtained from the CL(2) and granularity for parameters $\beta_j, \sigma_j$ are close, although slight asymmetries arise in the histograms of some of these parameters estimated by the granularity.

In addition, the granularity approach is computationally more intensive, and the computation burden increases with the number of factor values, given that the factor is estimated for each time period at the first step of the granularity estimation.

{\renewcommand{\arraystretch}{1.2}
{\setlength{\tabcolsep}{0.2cm} 
\begin{table}[ptbh]
\caption{Average Estimation Time (in seconds) of Parameters}%
\label{Computation}%
\bigskip
\centering%
\resizebox{\columnwidth}{!}{%
\begin{tabular}
[c]{l|l|l|l|l|l|l|l|l|l|l|l|l}\hline\hline
& \multicolumn{4}{|l}{CL(1)} & \multicolumn{4}{|l}{CL(2)} & \multicolumn{4}{|l}{Granularity Approach}%
\\\hline
& $\rho=0.0$ & $\rho=0.4$ & $\rho=0.7$ & $\rho=0.95$ &$\rho=0.0$  & $\rho=0.4$ & $\rho=0.7$ & $\rho=0.95$ &$\rho=0.0$  & $\rho=0.4$ & $\rho=0.7$ & $\rho=0.95$\\\hline\hline
$T=60$& 13.46& 14.06 & 15.42& 22.88 &   105.08 & 103.74  & 105.64& 103.19   &74.57 & 74.20 &73.95& 73.70  \\
$T=120$ &12.75  & 13.20 & 14.17& 19.25 &  105.74 & 104.10 & 105.54& 102.77 & 153.54 &  153.40 & 150.04& 149.93 \\
$T=240$ & 11.53 & 12.01 & 11.99& 15.67 &   104.97&104.26  & 105.19& 103.04 &  294.93 & 299.92 & 286.38 & 285.32 \\\hline\hline
\end{tabular}

}
\end{table}}
To give an idea of the computation time, the average estimation times are presented in Table \ref{Computation}. The estimation is carried out on a virtual machine running in a VMware (Virtual Machine) cluster.\footnote{Intel Xeon Gold 6140, multi threaded CPU with 20 real cores, 192GB ECC RAM, and 1TB enterprise grade SAS drive disk space with RAID-6, and dual power supply.} The code is written in Matlab and the optimizations are performed by using the fminsearch.
 The results show that the maximum CL(1) likelihood estimation is at least  $5$ up to $7$ times faster than the CL(2) estimation and granularity approach when $T=120$, respectively. Furthermore, it is at least  $6$ and $18$  times faster than the CL(2) estimation and granularity approach when $T=240$. Overall, the granularity procedure is more intensive (and requires maximizations with respect to higher numbers of parameters)  than the CL(1) and the CL(2) methods as the number of time periods increase, while the CL(1) is always the fastest in terms of computation. 

\section{Empirical Study}
In this section, we present the empirical results. Subsection 6.1 introduces the data. In Section 6.2, we estimate the model from the conditional composite log-likelihood at lag one proposed in Section 3.2, given that CL(1) performs well in the finite sample experiments and is computationally less intensive. We analyze the estimated parameters, transition probabilities, probabilities of defaults, and the downgrade probabilities at different horizons.

\subsection{Data Description}

The observed transition probabilities are computed from the Compustat S\&P rating database over the period 1985Q4  to 2016Q4. The average number of firms over all the $125$ quarters is $13,599$. In this section, we describe the data set and explain the necessary transformation due to missing data on non-rated companies.

We use the domestic long-term issuer quarterly credit ratings classified in eight categories: AAA, AA, A, BBB, BB, B, CCC/CC, and D, ranked from the lowest up to the highest risk, as in Section 5: $k=1,\ldots,8$, respectively. Each transition matrix summarizes all rating movements across these categories over one quarter.

As explained by \cite{Feng2008}, non-rated firms arise when the relevant debt is extinguished and there is a lack of balance sheet information to determine the firm rating due to a merger or an acquisition.
We follow the approach of \cite{Feng2008} to correct for non-rated firms. More precisely, we use for our analysis the transition probability conditional on being rated at the end of the quarter. We divide the frequency of migrating from any rating $j=1,\ldots,7$ to $k=1,\ldots,8$ by one minus the frequency of migrating from $j=1,\ldots,7$ to the non-rated state $j=9$. The resulting non-rated adjusted transition matrix 
is presented in Table \ref{Emp_2}. 

{\renewcommand{\arraystretch}{0.9}
\begin{table}[ptbh]
\caption{Non-Rated Adjusted Transition Matrix for 1987Q2 (in \%)}%
\label{Emp_2}
\bigskip
\centering%
%
\begin{tabular}
[c]{c|c|c|c|c|c|c|c|c}\hline\hline
&$k=1$ & $k=2$ & $k=3$ & $k=4$ & $k=5$ & $k=6$ & $k=7$ & $k=8$ \\\hline\hline
$j=1$ & 94.44 & 2.78 & 0.00 & 2.78 & 0.00 & 0.00 & 0.00 & 0.00 \\
$j=2$ & 0.00 & 99.03 & 0.97 & 0.00 & 0.00 & 0.00 & 0.00 & 0.00 \\
$j=3$  & 0.00 & 0.58 & 97.10 & 1.45 & 0.29 & 0.58 & 0.00 & 0.00 \\
$j=4$  & 0.00 & 0.81 & 1.63 & 92.68 & 3.66 & 1.22 & 0.00 & 0.00 \\
$j=5$ & 0.00 & 0.00 & 0.00 & 2.44 & 96.10 & 0.98 & 0.00 & 0.48 \\
$j=6$  & 0.00 & 0.00  & 0.00  & 0.37 & 1.87 & 95.90 & 0.75 & 1.11 \\
$j=7$  & 0.00 & 0.00  & 0.00  & 0.00 & 2.33 & 0.00 & 95.35 & 2.32 \\\hline\hline
\end{tabular}
\end{table}}

The frequencies of firms remaining in the same ratings over 1987Q2 is close to $100\%$ because the changes of ratings do not occur very often. The next highest transition probabilities are on the two diagonals below and above the 
main diagonal, as firms generally move away by one category from their initial rating if they are up-graded or down-graded by the rating agencies. The last column shows the probability that a rated firm defaults. The migration matrices in the sample have a similar pattern to the one documented in Table \ref{Emp_2}.

The S$\&$P database provides the rating structure of issuers at each date (see the first column of Table \ref{Emp_2}). This structure changes over time due to rating migration, and also because of defaults of some issuers and the arrivals of new issuers. Therefore, there is another type of rebalancing of the population of firms. We use the time averages of these structures to define the weights in the estimation.

\subsection{Empirical Results}

We report in Table \ref{CL1}, the estimated parameters and their bootstrap confidence intervals.\footnote{The firms are randomly drawn with replacements. Consequently, the firms' histories are kept unchanged once a firm is drawn. Therefore, we estimate parameters from the bootstrap data and use them to find the bootstrap intervals of the estimated parameters based on $B=399$ replications.} Table \ref{ETM1} contains the estimated transition matrix, and Table \ref{DPD} presents some resulting downgrade probabilities and probabilities of default.

{\renewcommand{\arraystretch}{1.0}
{\setlength{\tabcolsep}{0.3cm} 
\begin{table}[ptbh]
\caption{Estimated Parameters}%
\label{CL1}
\bigskip
\centering%
\resizebox{\columnwidth}{!}{%
\begin{tabular}
[c]{cccccccc}\hline\hline
&$j=1$ & $j=2$ & $j=3$ & $j=4$ & $j=5$ & $j=6$ & $j=7$  \\\hline\hline
$c_{j+1}$ &  & 6.18 & 6.40 & 9.01 & 11.05 & 15.20 & 17.09  \\
       &  & (4.05, 6.74) & (4.37, 6.83) & (5.90,  9.16)  & (7.86, 11.53) & (12.89, 18.94) & (14.44, 23.05) \\\hline
$\delta_j$ &-1.92 & 4.75 & 6.30 & 7.76 & 10.06 & 13.17 & 16.13  \\
       & (-2.95, 0.70 ) & (2.75, 5.87) & (4.23, 6.78) & (5.30, 7.83) & ( 7.02, 0.11) & (11.13, 14.97) & (13.81, 21.03)  \\\hline
$\gamma_j$  &  &   0.71 & 0.04 & 0.59 & 0.52 & 1.05 & 0.76  \\
   &  & (0.26, 1.12) & (0.01, 0.11)  & (0.07, 0.65) & (0.22, 0.93) & (0.78, 2.52) & (0.43, 1.83)  \\\hline\hline
\end{tabular}
}
\end{table}}}
     
The estimated thresholds in Table \ref{CL1} increase with the ratings, so that firms with higher latent scores receive higher ratings. Furthermore, firms with higher ratings have higher estimated intercepts and therefore increased scores. The largest estimated value of unconditional volatility is obtained for $j=6$ and $j=7$, which corresponds to firms facing major uncertainties, currently vulnerable, or which have filed for bankruptcy protection. 

In practice, the parameters of interest can be nonlinear functions of $c_{j}$, $\delta_j$ and $\gamma_j$. For instance, one might be interested in the quasi-transition probabilities, the quasi-downgrade probabilities and the quasi-probabilities of default. We used the estimates to compute the migration probabilities and illustrate the prediction of downgrade probabilities and probabilities of default as follows.
\begin{description}
\item[i)] the downgrade probabilities at horizon 1 and 2 of a firm currently rated $j$: $DP(1|j),DP(2|j)$.

\item[ii)] the term structure of the probability of default at different
horizons $h$ for a firm currently rated $j$. The horizons are fixed to
$1$ quarter, $3$ year, $6$ years, $9$ years, and denoted by $PD(1|j),PD(12|j),PD(24|j),PD(36|j)$.
\end{description}

{\renewcommand{\arraystretch}{0.9}
\begin{table}[ptbh]
\caption{Estimated Quasi-Migration Matrix (in \%)}%
\label{ETM1}
\bigskip
\centering%
%
\begin{tabular}
[c]{c|c|c|c|c|c|c|c|c}\hline\hline
   &$k=1$ & $k=2$ & $k=3$ & $k=4$ & $k=5$ & $k=6$ & $k=7$ & $k=8$ \\\hline\hline
$j=1$ & 97.28 & 2.72 & 0.00 & 0.00 & 0.00 & 0.00  & 0.00 & 0.00	 \\
$j=2$ & 0.00 & 97.81 & 1.20 & 0.99 & 0.00 & 0.00 & 0.00 & 0.00	 \\
$j=3$ & 0.00 & 0.20 & 98.09 & 1.71 & 0.00 & 0.00 & 0.00 & 0.00 \\
$j=4$  &  0.00 & 0.39 & 0.72 & 97.13 & 1.76 & 0.00 & 0.00 & 0.00\\
$j=5$  &0.00 & 0.00 & 0.00 & 2.12 & 95.14 & 2.74 & 0.00 & 0.00 \\
$j=6$ & 0.00 & 0.00 & 0.00 & 0.00 & 2.21 & 95.12 & 2.66 & 0.01 \\
$j=7$  & 0.00 & 0.00 & 0.00 & 0.00 & 0.00 & 10.94 & 78.80 & 10.26 \\\hline
\end{tabular}
\end{table}}
\bigskip

Table \ref{ETM1} shows the estimates of the expected migration probabilities from any state $j=1,\ldots,7$ to 
$k=1,\ldots,8$, after plugging in  the parameter estimates. The estimated quasi-migration matrix reproduced the  aforementioned features of 
an observed transition matrix. Table \ref{DPD} presents the downgrade probabilities and the probabilities of default 
computed from the estimated transition matrix. We observe that the downgrade probability and the probability of 
default often increase as the horizon increases. At horizon 1, the downgrade probability is $2.72\%$ for a firm initially rated 
$j=1$, while it increases to $10.26\%$ for a firm initially rated $j=7$. At horizon 2, the estimated probability increases 
to $18.35\%$ for $j=7$ from $5.37$ for $j=1$. At horizon $1$, the probabilities of default are zero for 
ratings $j=1,\ldots,5$. This is consistent with the fact that the probability that a firm with a better capacity to repay its 
debt defaults during one quarter is negligible. The estimated probabilities increase as the horizon increases.

{\renewcommand{\arraystretch}{1.0}
\begin{table}[ptbh!]
\caption{Estimated Downgrade Probabilities and Probabilities of Default (in \%)}%
\label{DPD}%
\bigskip\centering%

\begin{tabular}{cccccccc}
\hline\hline
  & \multicolumn{1}{|c}{$j=1$} & \multicolumn{1}{|c}{$j=2$} & 
\multicolumn{1}{|c}{$j=3$} & \multicolumn{1}{|c}{$j=4$} & \multicolumn{1}{|c}{$j=5$} & \multicolumn{1}{|c}{$j=6$} & \multicolumn{1}{|c}{$j=7$} \\ \hline\hline
$DP\left( 1|j\right) $ & \multicolumn{1}{|c}{$2.72$} & \multicolumn{1}{|c}{$2.19$} & \multicolumn{1}{|c}{$1.71$} & \multicolumn{1}{|c}{$1.76$}& \multicolumn{1}{|c}{$2.75$} & \multicolumn{1}{|c}{$2.67$} & \multicolumn{1}{|c}{$10.26$}  \\ 
$DP\left( 2|j\right) $ & \multicolumn{1}{|c}{$5.37$} & \multicolumn{1}{|c}{$%
4.32$} & \multicolumn{1}{|c}{$3.37$} & \multicolumn{1}{|c}{$3.44$} & \multicolumn{1}{|c}{$5.30$} & \multicolumn{1}{|c}{$4.91$} & \multicolumn{1}{|c}{$18.35$} \\ \hline

$PD\left( 1|j\right) $ & \multicolumn{1}{|c}{$0.00$} & \multicolumn{1}{|c}{%
$0.00$} & \multicolumn{1}{|c}{$0.00$} & \multicolumn{1}{|c}{$0.00$}& \multicolumn{1}{|c}{$0.00$} & \multicolumn{1}{|c}{$0.01$} & \multicolumn{1}{|c}{$10.26$}  \\ 

$PD\left( 12|j\right) $ & \multicolumn{1}{|c}{$0.00$} & \multicolumn{1}{|c}{$%
0.00$} & \multicolumn{1}{|c}{$0.00$} & \multicolumn{1}{|c}{$0.04$}& \multicolumn{1}{|c}{$0.86$} & \multicolumn{1}{|c}{$8.33$} & \multicolumn{1}{|c}{$48.00$} \\ 

$PD\left( 24|j\right) $ & \multicolumn{1}{|c}{$0.00$} & \multicolumn{1}{|c}{$0.02$} & \multicolumn{1}{|c}{$ 0.04 $} & \multicolumn{1}{|c}{$0.43$}& \multicolumn{1}{|c}{$3.93$} & \multicolumn{1}{|c}{$18.01$} & \multicolumn{1}{|c}{$55.68$}  \\ 
$PD\left( 36|j\right) $ &  \multicolumn{1}{|c}{$0.02$} & \multicolumn{1}{|c}{$0.12$} & \multicolumn{1}{|c}{$0.18$} & \multicolumn{1}{|c}{$1.37$}& \multicolumn{1}{|c}{$7.91$} & \multicolumn{1}{|c}{$25.43$} & \multicolumn{1}{|c}{$60.07$} \\ \hline\hline
\end{tabular}
\end{table}
}

\section{Conclusion}
The stochastic factor ordered Probit model has been introduced for dynamic analysis of credit risk, as it is sufficiently flexible to account  for the rating dynamics, the presence of systemic risk and the stylized fact of rating momentum. 
This paper proposes three maximum composite likelihood estimation methods of different complexity for this model: the conditional composite log-likelihood function at lag 1, the conditional composite log-likelihood at lag 2, and the conditional composite likelihood up to lag 2. The paper discusses the identifiability of the model parameters and establishes the asymptotic properties of these estimators when both the cross-sectional dimension $n$ and the time dimension $T$ tend to infinity. In this asymptotic setup, the MCL methods are less efficient than the two-step granularity-based approach existing in the literature. However, in practice $n$ is large, but $T$ is much smaller. 
The considerable advantage of the MCL approaches is that they reduce the computational burden of the granularity approach as they avoid estimating a large number of nuisance parameters. 
We illustrate the finite sample properties of the conditional composite log-likelihood at lag 1 and  at lag 2 by conducting Monte-Carlo experiments and compare with the granularity approach. Our results indicate that the MCL methods are reliable at T=60 and are computationally less demanding than the granularity-based estimator at finite $T$. An empirical study illustrates the application of the proposed method to credit rating data.

\begin{appendices}
\setcounter{secnumdepth}{0}
\section{Appendix A: The Expected Transition Matrices}

\subsection{A.1. Expected Matrix $P$ (Lemma 1)}
\indent The matrix $P$ is computed from
$$ y_{i,t}^{*}=\beta_{j}f_{t}+\delta_{j}+\sigma_{j}u_{i,t},\; \mbox{if} \;\;
y_{i, t-1} = j, $$

\noindent as if $u_{i,t}\sim N(0,1)$ and $f_{t}\sim N(0,1)$ were independent. Then, under this independence condition, if $y_{i,t-1}=j$, $y_{i,t}^*|y_{i,t-1}=j\sim N(\delta_{j}, \sigma_{j}^{2}+\beta_{j}^{2})$.
\noindent It follows that
\begin{align*}
P[y_{i,t}=k|y_{i,t-1}=j]&=P[c_{k}<y^{*}_{i,t}<c_{k+1}|y_{i,t-1}=j],
\end{align*}
and
\begin{align*}
 p_{jk}(\theta) &=\Phi\left(\frac{c_{k+1}-\delta_{j}}{\sqrt{\sigma_{j}^{2}+\beta_{j}^{2}}}\right)-\Phi\left(\frac{c_{k}-\delta_{j}}{\sqrt{\sigma_{j}^{2}+\beta_{j}^{2}}}\right).
 \end{align*}

\subsection{A.2. Matrix $P^{(2)}$ (Lemma 2)}
\indent We have
\begin{align*}
    P^{(2)}&=E[P(f_{t};\theta) \  P(f_{t-1};\theta)]=E[P(\rho f_{t-1}+\sqrt{1-\rho^{2}} \ \eta_{t};\theta)] \  P(f_{t-1};\theta)].
\end{align*}

\noindent where the expectation is taken with respect to the joint marginal distribution of $(f_t, f_{t-1})$.
Since $f_{t-1}$ and $\eta_{t}$ are independent, $\eta_{t} \sim N(0,1)$ and $f_{t-1}\sim N(0,1)$, we get
\begin{align*}
P^{(2)} &= E_{f_{t-1}} \ E_{\eta_{t}} \bigg[ P(\rho f_{t-1}+\sqrt{1-\rho^{2}} \ \eta_{t};\theta) \  P(f_{t-1};\theta)|f_{t-1}\bigg],\\
&= E_{f_{t-1}} \bigg[ E_{\eta_{t}} \bigg[ P(\rho f_{t-1}+\sqrt{1-\rho^{2}} \ \eta_{t};\theta)|f_{t-1}\bigg] \  P(f_{t-1};\theta)\bigg],\\
&=E_{f_{t-1}} \bigg[ A \ B\bigg],
\end{align*}
where the components of matrix $A$ are given by
\begin{align*}
a_{kl} (f_{t-1}; \theta, \rho) &=\mathbb{P}\bigg[ c_{k}<y_{i,t}^{*}<c_{k+1}| \ y_{i,t-1}=l,f_{t-1}\bigg]\\
&=  \mathbb{P} \bigg[ c_{k}<\delta_{l}+\beta_{l}\rho f_{t-1}+\beta_{l}\sqrt{1-\rho^{2}} \ \eta_{t}+\sigma_{l}u_{i,t}<c_{k+1}|  f_{t-1} \bigg], \\
&=\Phi\left(\frac{c_{k+1}-\delta_{l}-\beta_{l}\rho f_{t-1}}{\sqrt{\sigma_{l}^{2}+\beta_{l}^{2}(1-\rho^{2})}}\right)-\Phi\left(\frac{c_{k}-\delta_{l}-\beta_{l}\rho f_{t-1}}{\sqrt{\sigma_{l}^{2}+\beta_{l}^{2}(1-\rho^{2})}}\right), k, l, =1,...K,
\end{align*}
as  if  $\left(\eta_t,u_{i,t}\right)$ and $\left(y_{i,t-1},f_{t-1}\right)$ were independent. By (2.4) the elements of matrix $B$ are
\begin{align*}
p_{jl}(f_{t-1};\theta)=\Phi\left(\frac{c_{l+1}-\delta_{j}-\beta_{j} f_{t-1}}{\sigma_{j}}\right)-\Phi\left(\frac{c_{l}-\delta_{j}-\beta_{j} f_{t-1}}{\sigma_{j}}\right), j, l=1,...,K.
\end{align*}
\noindent Therefore, by integrating out $f= f_{t-1}$, we get
\begin{align*}
p_{jk}^{(2)}(\theta,\rho)&=\int \ \sum_{l=1}^{K} [a_{k,l}(f;\theta, \rho) \ p_{j,l}(f;\theta)] \phi (f)df\\
&=\int \ \sum_{l=1}^{K} \bigg[\Phi\left(\frac{c_{k+1}-\delta_{l}-\beta_{l}\rho f}{\sqrt{\sigma_{l}^{2}+\beta_{l}^{2}(1-\rho^{2})}}\right)-\Phi\left(\frac{c_{k}-\delta_{l}-\beta_{l}\rho f}{\sqrt{\sigma_{l}^{2}+\beta_{l}^{2}(1-\rho^{2})}}\right) \bigg]\\
&\times\bigg[ \Phi\left(\frac{c_{l+1}-\delta_{j}-\beta_{j} f}{\sigma_{j}}\right)-\Phi\left(\frac{c_{l}-\delta_{j}-\beta_{j} f}{\sigma_{j}}\right)\bigg]\phi(f) df. 
\end{align*}

\section{Appendix B: Proof of Propositions 1 and 2}
\subsection{B.1. Proof of Proposition 1}
\indent From Lemma 1 and the definitions $c_1=-\infty$, $c_{K+1}=\infty$, we know that the identifiable functions of parameters are
\begin{align*}
\frac{c_{k}-\delta_{j}}{\sqrt{\sigma_{j}^{2}+\beta_{j}^{2}}} \ \ \  \forall  \ k=2, ..., K, \; j=1, ..., K. \;\;\;\;\;\;\; \;\;\;\;\;\;\; \;\;\;\;\;\;\;\text{(b.1)}
\end{align*}
Therefore parameter $\rho$ is not identifiable. Moreover, parameters $\sigma_{j}^{2}$ cannot be distinguished from $\beta_{j}^{2}$. Let us denote their sum by $\gamma_{j}^2$, where
\begin{align*}
\gamma_{j}=\sqrt{\sigma_{j}^{2}+\beta_{j}^{2}}.    
\end{align*}
\noindent There are $K(K-1)$ identifying functions (b.1), that we would like to use to identify the $(K-1)$ values of $c_{k}$, the $K$ values of $\delta_{j}$ and the $K$ values of $\gamma_{j}$, i.e. $3K-1$ unknowns. We follow \cite{GG2015} and add the identifying restrictions:
\begin{align*}
    c_{2}=0, \ \ \gamma_{1}=1.
\end{align*}
\noindent Next, we proceed as follows.

\noindent (a) \ From (b.1) written for $k=2$, we identify $\cfrac{\delta_{j}}{\gamma_{j}}$. Given that $\gamma_{1}=1$, we get $\delta_{1}$ identified.

\noindent (b) \ For $j=1$, we have $\gamma_{1}=1$, hence we identify $c_{k}-\delta_{1}$, given (b.1). Therefore, all thresholds $c_{k}, \ \ k=2, ..., K$ are identified.

\noindent (c) \ Then the  identifying functions can also be written as
\begin{align*}
    \frac{c_{k}-\delta_{j}}{\gamma_{j}}= \frac{c_{k}}{\gamma_{j}}-\frac{\delta_{j}}{\gamma_{j}}, k=2,\ldots,K, j=1,\ldots,K.
\end{align*}
Therefore, from the identification of the ratios $\delta_j/\gamma_j$ result in (a), we identify all ratios $c_k/\gamma_j$.
Then from the identification of the $c_k$'s (b), we identify $\gamma_{j}$, $j=1, ..., K$. Next, the $c_k, \gamma_j$ are identified, and from (c), we identify $\delta_{j}$, $j=1, ..., K$. 

\subsection{B.2. Proof of Proposition 2}
\indent We have the following identifying functions of parameters
\begin{align*}
&(1) \ \ \frac{c_{k}-\delta_{j}}{\sqrt{\sigma_{j}^{2}+\beta_{j}^{2}(1-\rho^{2})}}, k=2,\ldots,K, j=1,\ldots,K\\
&(2) \ \ \frac{\epsilon \beta_{j}\rho}{\sqrt{\sigma_{j}^{2}+\beta_{j}^{2}(1-\rho^{2})}}, j=1,\ldots,K\\
&(3) \ \ \frac{c_{k}-\delta_{j}}{\sigma_{j}}, k=2,\ldots,K, j=1,\ldots,K\\
&(4) \ \ \frac{\epsilon \beta_{j}}{\sigma_{j}}, j=1,\ldots,K.
\end{align*}
\noindent Let us define
\begin{align*}
\gamma_{j}=\sqrt{\sigma_{j}^{2}+\beta_{j}^{2}(1-\rho^{2})},
\end{align*}
\noindent and use the identifying restrictions
\begin{align*}
\gamma_{1}=1, \ \ c_{2}=0. 
\end{align*}
\noindent Then we proceed as follows.

\noindent (a) \ For $k=2$, given $c_{2}=0$,

\indent and identified function (1), we identify $\delta_{j}/\gamma_{j}$. 

\noindent (b) \ For $k=2$, given $c_{2}=0$,

\indent and function (3), we identify $\delta_{j}/\sigma_{j}, j=1,\ldots,K$.

\noindent (c) \ Given that $\gamma_{1}=1$, it follows from (a) that parameter $\delta_{1}$ is identified.

\noindent (d) \ Then, it follows from (b) that parameter $\sigma_{1}$ is identified.

\noindent (e) \ For $j=1$ and identified function (1), we identify:
\begin{align*}
    \frac{c_{k}-\delta_{1}}{\gamma_{1}}=c_{k} - \delta_1.
\end{align*}
Hence, from (c), it follows that $c_{k}$, $k=1, ..., K-1$ are identified.

\noindent (f) \ From identified function (1), the quantities
\begin{align*}
\frac{c_{k}}{\gamma_{j}}-\frac{\delta_{j}}{\gamma_{j}},
\end{align*}
are identified  since $\gamma_1=1$.

\noindent Then,  by (a), the ratios $c_{k}/\gamma_{j}$ are identified. 

\noindent (g) \ From (f) and (e), parameters $\gamma_{j}$, $j=1, ..., K$ are identified.

\noindent (h) \ From (a) and (g), parameters $\delta_{j}$, $j=1, ..., K$ are identified. 

\noindent (i) \ From (b) and (h), parameters $\sigma_{j}$, $j=1, ..., K$ are identified. 

\noindent (j) \ From equation (4) and result (i), parameters $\epsilon \beta_{j}$, $j=1, ..., K$ are identified.

\noindent (k) \ From (2), we get the ratios $\epsilon \beta_{j} \rho/\gamma_{j}$ and given (g) we identify $\epsilon \beta_{j}\rho$, $j=1, ..., K$.

\noindent (l) \ Finally, from (j) and (k), we identify parameter $\rho$. 

\section{Appendix C: Proof of Uniform a.s. Convergence}

Let us introduce a more precise notation: $\hat{p}_{jk,t}(n)$, where the argument $n$ is introduced to describe the dependence of the transition frequencies on the number of individuals $n$,
and consider the assumption A.4 i),

$$
P [ Max_{m \geq n} | \hat{p}_{jk,t}(m) - p_{jk}(f_t, \theta_0)| > \epsilon|f_t] < \frac{1}{\epsilon^2 n} g_{jk}(f_t, \theta_0), \;\forall \epsilon >0, \, \forall j,k,f_t,
$$

\noindent Then, it follows that
$$P [ Max_{m \geq n} |\hat{p}_{jk,t}(m) - p_{jk}(f_t, \theta_0)| > \epsilon|f_t] < \frac{1}{\epsilon^2 n} \sum_{t=2}^T g_{jk}(f_t, \theta_0), \, \forall j,k,f_t.$$
\noindent For $n$ large, the upper bound: $\frac{1}{\epsilon^2}\frac{T}{n} \frac{1}{T} \sum_{t=2}^T g_{jk}(f_t, \theta)$
is equivalent to $\frac{1}{\epsilon^2} \frac{T}{n} E_0 [ g_{jk}(f_t, \theta_0)]$, by the geometric ergodicity of factor $(f_t)$. Then by Assumption A4 ii),  $T \rightarrow \infty, n  \rightarrow \infty$ with $T/n \rightarrow 0$, we infer
$$\lim_{n \rightarrow \infty, T \rightarrow \infty} Sup_{t \leq T} P [ Max_{m \geq n} | \hat{p}_{jk,t}(m) - g_{jk}(f_t, \theta_0)| > \epsilon| f_t] =0,$$
and the required uniformity .

\noindent Therefore, after the normalization, the a.s. limit of the normalized composite log-likelihood is
\begin{align*}
& \lim_{n,T\rightarrow{\infty}} \mbox{a.s.} \ \frac{1}{T} \sum_{k=1}^{K}\sum_{j=1}^{K}\sum_{t=2}^{T}\bigg[ \pi_j \ \hat{p}_{jk,t}(n) \ log \ p_{jk}(\theta)\bigg] \\ 
=& \lim_{T\rightarrow{\infty}} \mbox{a.s.} \ \frac{1}{T} \sum_{l=1}^{K}\sum_{k=1}^{K}\sum_{t=2}^{T}\bigg[ \pi_j \lim_{n\rightarrow{\infty}} \hat{p}_{jk,t}(n) \ log \ p_{jk}(\theta)\bigg]\\
=&\lim_{T\rightarrow{\infty}} \mbox{a.s.} \ \frac{1}{T} \sum_{l=1}^{K}\sum_{k=1}^{K}\sum_{t=2}^{T}\Bigg[ \pi_j p_{jk}(f_t, \theta_0) \ log \ p_{jk}(\theta)\Bigg]\\
&\mbox{(by the uniform a.s. convergence)}\\
=&\sum_{j=1}^{K}\sum_{k=1}^{K}\Bigg[ \pi_j \lim_{T\rightarrow{\infty}} \mbox{a.s.}\bigg[ \frac{1}{T} \sum_{t=2}^{T} p_{jk}(f_t, \theta_0) \log \ p_{jk}(\theta)\bigg]\Bigg]\\     
=&\sum_{j=1}^{K}\Bigg[ \ \pi_j \ \bigg[\sum_{k=1}^{K} p_{jk}(\theta_{0}) \ log \ p_{jk}(\theta)\bigg]\Bigg] \mbox{(since $f_t$ is geometrically ergodic)}.
\end{align*}
Therefore, $L_{cc} ( \theta)$ converges a.s. uniformly to
$$L_{cc}^\infty (c, \delta, \gamma)= \sum_{j=1}^{K}\Bigg[ \pi_j \ \bigg[\sum_{k=1}^{K} p_{jk}(\theta_{0}) \ log \ p_{jk}(\theta)\bigg]\Bigg]. $$

\end{appendices}


\bigskip

 \begin{thebibliography}{999999999999999999999999999999999999999999}   
 
 \bibitem[Altman, Kao (1992):]{AK1992}\textsc{Altman, E. I.}, and \textsc{D. L., Kao} (1992).  \textit{Rating Drift in High-Yield Bonds}. The Journal of Fixed Income, 1, 15--20.
 
\bibitem[Altman, Saunders (1998):]{altman1997credit}\textsc{Altman, E. I.}, and \textsc{A., Saunders} (1998).  \textit{Credit Risk Measurement: Developments Over the Last 20 Years}. Journal of Banking and Finance, 21, 1721--1742.

 \bibitem[Azizpour, Giesecke, Schwenkler (2018): ]{AGS2018} \textsc{Azizpour, S.}, \textsc{ Giesecke, K.}, and \textsc{G., Schwenkler} (2018).  \textit{Exploring the Sources of Default Clustering}, Journal of Financial Economics, 129, 154--183.
 
\bibitem[Bangia, Diebold, Kronimus, Schlagen, and Schuerman (2002):]{BDKSS2002}  \textsc{Bangia, A.}, \textsc{Diebold, F.}, \textsc{Kronimus, A}., \textsc{Schlagen, C.}, and \textsc{T. Schuerman} (2002). \textit{Rating Migration and the Business Cycle with Application to Credit Portfolio Stress Testing}, Journal of Banking and Finance, 26, 445--474.
 
\bibitem[Basel Committee on Banking Supervision (2004): ]{bas2} \textsc{Basel Committee on Banking Supervision} (2004).  \textit{International Convergence of Capital Measurement and Capital Standards: A Revised Framework}, Technical report, Bank for International
Settlements.

\bibitem[Basel Committee on Banking Supervision (2009): ]{bas3} \textsc{Basel Committee on Banking Supervision} (2009).  \textit{Guidelines for Computing Capital for Incremental Risk in the Trading Book}, Technical report, Bank for International
Settlements.

\bibitem[Berndt, Douglas, Duffie, Fergusson (2018): ]{BDDF2018} \textsc{Berndt, A.,} \textsc{Douglas, R.,} \textsc{Duffie, D.,}  and \textsc{M., Fergusson} (2018).  \textit{Corporate Credit Risk Premia}, Review of Finance, 22, 419--454.

\bibitem[Billingsley (1961):]{Billingsley1961} \textsc{Billingsley, P.} (1961): \textsc{Statistical Inference for Markov Processes}, The University of Chicago Press.

 \bibitem[Bonhomme, Jochmans, Robin (2017):]{BJR2017} \textsc{Bonhomme, S.}, \textsc{Jochmans, K.}, and \textsc{J.M., Robin} (2017). \textit{Nonparametric Estimation of Non-Exchangeable Latent-Variable Models}, Journal of Econometrics, 201, 237--248.

 \bibitem[Cousin, Lelong, Picard (2021):]{CLP2021} \textsc{Cousin A.}, \textsc{Lelong J.,} and \textsc{T. Picard} (2022). \textit{Rating Transitions Forecasting: a Filtering Approach}, Papers 2109.10567, arXiv.org.

\bibitem[Cox, Reid (2004): ]{CR2004}\textsc{Cox, D.}, and \textsc{N., Reid} (2004). \textit{A Note on Pseudolikelihood Constructed from Marginal Densities}. Biometrika, 91, 729-737.

\bibitem[Creal, Koopman, Lucas (2012):]{Creal2013} \textsc{Creal, D.}, \textsc{Koopman, S.}, and \textsc{A., Lucas} (2012).  \textit{Generalized Autoregressive Score Models with Applications}, Journal of Applied Econometrics, 28, 777--795.

\bibitem[Creal, Schwaab, Koopman, Lucas (2014):]{Creal2014} \textsc{Creal, D.}, \textsc{Schwaab, B.}, \textsc{Koopman, S.}, and \textsc{A., Lucas} (2014).  \textit{Observation-Driven Mixed-Measurement Dynamic Factor Models with an Application to Credit Risk}, Review of Economics and Statistics, 96, 898--915.

\bibitem[Crouhy, Galai, and Mark (2000):]{CGM2000} \textsc{Crouhy, M.}, \textsc{Galai, D.}, and \textsc{R. Mark} (2000). \textit{A Comparative Analysis of Current Credit Risk Models}, Journal of Banking and Finance, 29, 59--117.

\bibitem[Dos Reis, Pfeuffer, Smith (2020):]{DPS2020} \textsc{dos Reis, G.}, \textsc{ Pfeuffer, M.}, and \textsc{G., Smith} (2020).  \textit{Capturing Model Risk and Rating Momentum in the Estimation of Probabilities of Default and Credit Rating Migrations}, Quantitative Finance, 20, 1069--1083.

\bibitem[Duffie, Eckner, Horel, Saita (2009): ]{DEHS2009} \textsc{ Duffie, D.}, \textsc{ Eckner, A.}, \textsc{ Horel, G.}, and \textsc{L., Saita} (2009).  \textit{Frailty Correlated Default}, The Journal of Finance, 64,2089-- 2123.

\bibitem[European Banking Authority (2012):]{EBA2012} \textsc{European Banking Authority} (2012). \textit{Guidelines on the Incremental Default and Migration Risk Change (IRC)}, EBA/GL/2012/3.

\bibitem[Farmer (2021):]{Farmer2021} \textsc{Farmer, L.} (2021).  \textit{The Discretization Filter: A Simple Way to Estimate Nonlinear State Space Models}, Quantitative Economics, 12, 41--76.

\bibitem[Feng, Gouri\'eroux, Jasiak (2008) :]{Feng2008} \textsc{Feng, D.}, \textsc{ Gouri\'eroux, C.}, and \textsc{J., Jasiak} (2008). \textit{ The Ordered Qualitative Model for Credit Rating Transitions}, Journal of Empirical Finance, 15, 111--130.

\bibitem[Frydman, Schuermann (2008): ]{FS2008} \textsc{Frydman, H.}, and \textsc{T.,Schuermann}  (2008). \textit{Credit Rating Dynamics and Markov Mixture Models}, Journal of Banking and Finance, 32, 1062--1075.

\bibitem[Gagliardini, Gouri\'eroux (2005): ]{GG2005a} \textsc{Gagliardini, P.}, and \textsc{C., Gouri\'eroux}  (2005). \textit{Stochastic Migration Models with Application to Corporate Risk}, Journal of Financial Econometrics, 3, 188--226.

\bibitem[Gagliardini, Gouri\'eroux (2014):]{GG2014} \textsc{Gagliardini, P.}, and \textsc{C., Gouri\'eroux} (2014).  \textit{Efficiency in Large Dynamic Panel Models with Common Factors}, Econometric Theory, 30, 961--1020.

\bibitem[Gagliardini, Gouri\'eroux (2015):]{GG2015} \textsc{Gagliardini, P.}, and \textsc{C., Gouri\'eroux} (2015).  \textit{Granularity Theory with Applications  to  Finance  and  Insurance}, Cambridge  University  Press,  186 pages.

\bibitem[Gavalus, Syriopoulos (2014)]{GS2014} \textsc{Gavalus, D.} and \textsc{T. Syriopoulos} (2014). \textit{Bank Credit Risk Management and Rating Migration Analysis on the Business Cycle}, International Journal of  Financial Studies, 2, 122--143.

\bibitem[Godambe (1960)]{Godambe1960} \textsc{Godambe, V. P.} (1960). \textit{An Optimum Property of Regular Maximum Likelihood Estimation}, The Annals of Mathematical Statistics, 31, 1208--1211.

\bibitem[Gomes-Gonzalo, Kiefer (2009):]{GK2009} \textsc{Gomes-Gonzalo, J.}, and \textsc{N. Kiefer} (2009). \textit{Evidence of Non-Markovian Behaviour in the Process of Bank Rating Migration}, Cuadernos de Economia, 46, 33--50. 

\bibitem[Gordy, Lutkebohmert (2013):]{GL2013}  \textsc{Gordy, M.}, and  \textsc{E. Lutkebohmert} (2013).  \textit{Granularity Adjustment for Regulatory Capital Assessment}, International Journal of Central Banking, 9, 38--77.

\bibitem[Gouri\'eroux, Monfort (2018):]{GM2018} \textsc{Gouri\'eroux, C.}, and \textsc{A., Monfort} (2018).  \textit{Composite Indirect Inference with Application to Corporate Risks}, Econometrics and Statistics, 7, 30--45.

\bibitem[Gouri\'eroux, Monfort, Polimenis (2006): ]{GMP2006} \textsc{Gouri\'eroux , C., } \textsc{Monfort, A., } and \textsc{V., Polimenis} (2006).  \textit{Affine Models for Credit Risk Analysis}, Journal of Financial Econometrics, 4, 494-530.

\bibitem[Gouri\'eroux, Monfort, Mouabbi, Renne (2021): ]{GMMR2021} \textsc{Gouri\'eroux , C., } \textsc{Monfort, A., } \textsc{Mouabbi, S., }  and \textsc{J.P., Renne} (2021).  \textit{Disastrous Defaults}, Review of Finance, 25, 1727-1772.

\bibitem[Grippa, Gornicka (2016):]{GG2016}  \textsc{Grippa, P.}, and  \textsc{L. Gornicka (2016)}.  \textit{Measuring Concentration Risk, A Partial Portfolio Approach}, IMF Working Paper 158. 

\bibitem[Harko, Lobo, Mak (2016): ]{HLM2014} \textsc{Harko, T.}, \textsc{Lobo, F.}, and \textsc{M., Mak} (2016).  \textit{Exact Analytical Solutions of the Susceptible-Infected-Recovered (SIR) Epidemic Model and of the SIR Model with Equal Death and Birth Rates}, Applied Mathematics and Computation, 236, 184--194.

\bibitem[Hirk, Vana, Hornik (2022):]{HVH2022} \textsc{Hirk, R.}, \textsc{Vana, L.}, and \textsc{K., Hornik} (2022). \textit{A corporate credit rating model with autoregressive errors}, Journal of Empirical Finance, 69, 224--240.

\bibitem[Huajian, Zunwei (2015):]{HZ2015} \textsc{Huajian, B.}, and \textsc{D. Zunwei} (2015). \textit{Stress Testing and Modelling of Rating Migration under the Vasicek Model Framework, Empirical Approaches and Technical Implementation}, Journal of Risk Model Validation, 9.

\bibitem[Hubert (1967): ]{Hubert1967} \textsc{Hubert, P.}(1967).  \textit{The Behavior of Maximum Likelihood Estimation under Nonstandard Conditions},  Proceedings of the Fifth Berkeley Symposium on Mathematical Statistics and Probability, Le Cam, L., and Neyman, J, eds, University of California Press, 221--233.

\bibitem[Hull (2012):]{hull2012risk} \textsc{Hull, J.} (2012). \textit{Risk Management and Financial Institutions}, John Wiley \& Sons, 733 pages.

\bibitem[Jennrich (1969): ]{Jennrich1969} \textsc{Jennrich, R. I.}(1969).  \textit{Asymptotic Properties of Non-Linear Least Squares Estimators}, Ann. Math. Statist., 40, 633--643.

\bibitem[Koopman, Lucas, Monteiro (2008): ]{Koopman2008} \textsc{Koopman, S.}, \textsc{Lucas, A.}, and \textsc{A., Monteiro}  (2008).  \textit{The Multi-State Latent Factor Intensity Model for Credit Rating Transitions}, Journal of Econometrics, 142, 399--424.

\bibitem[Lando, Sk\o deberg (2002): ]{LS2002} \textsc{Lando, D.}, and \textsc{T. M.,  Sk\o deberg}  (2002). \textit{Analyzing Rating Transitions and Rating Drift with Continuous Observations}, Journal of Banking and Finance, 26, 423--444.

\bibitem[Lindsay (1988): ]{Lindsay1988} \textsc{Lindsay, B. G.}(1988).  \textit{Composite Likelihood Methods}, Contemporary Mathematics, 80, 221--239.

\bibitem[Mizen, Tsoukas (2012):]{MT2012} \textsc{Mizen, P.}, and \textsc{S., Tsoukas} (2012). \textit{Forecasting US bond default ratings allowing for previous and initial state dependence in an ordered probit model}, International Journal of Forecasting, 28, 273--287.

\bibitem[Newey, West (1994):]{NW1994} \textsc{Newey, W.}, and \textsc{K., West} (1994).  \textit{Automatic Lag Selection in Covariance Matrix Estimation}, The Review of Economic Studies, 61(4), 631--653.

\bibitem[Nickell, Perraudin, Varotto (2000): ]{NPV2001} \textsc{Nickell, P.},  \textsc{Perraudin, W}, and \textsc{ S., Varotto } (2001). \textit{Stability of Rating Transitions}, Journal of Banking and Finance, 24, 203--227.

\bibitem[Reusens, Croux (2017): ]{RC2017} \textsc{Reusens, P.}, and \textsc{C., Croux} (2017).  \textit{Sovereign credit rating determinants: A comparison before and after
the european debt crisis}, Journal of Banking \& Finance, 77, 108--121.

\bibitem[Stefanescu, Tunaru, Turnbull (2009):]{STT2009} \textsc{Stefanescu, C.},  \textsc{Tunaru, R.},  and \textsc{S., Turnbull} (2009). \textit{The credit rating process and estimation of transition probabilities: A Bayesian approach}, Journal of Empirical Finance, 16, 216--234.

\bibitem[Tuzcuoglu (2022): ]{Tuzcuoglu2019} \textsc{Tuzcuoglu, K.} (2022).  \textit{Composite Likelihood Estimation of an Autoregressive Panel Probit Model with Random Effects}, Journal of Business and Economic Statistics, forthcoming.

\bibitem[Varian (2008): ]{Varian2008} \textsc{Varian, C.}(2008).  \textit{On Composite Marginal Likelihoods}, AStA Advances in Statistical Analysis, Springer; German Statistical Society, 92(1), 1--28.

\bibitem[Varian, Reid, Firth (2011): ]{VRF2011} \textsc{Varian, C.}, \textsc{Reid, N.}, and \textsc{D., Firth}  (2011).  \textit{An Overview of Composite Likelihood Methods}. Statistica Sinica, 5--42.

\bibitem[Vasicek (1991): ]{Vasicek1991} \textsc{Vasicek, O.} (1991).  \textit{Limiting Loan Loss Probability Distribution}. DP KMV Corporation.

\bibitem[Vasicek (2015): ]{Vasicek1987} \textsc{Vasicek, O.} (2015).  \textit{Probability Loss and Loan Portfolio}. In: Vasicek, O.A. (ed.) Finance, Economics and Mathematics. Hoboken, NJ: John Wiley \& Sons,
Inc., Chapter 17.

\bibitem[White (1982): ]{White1982} \textsc{White, H.} (1982).  \textit{Maximum Likelihood Estimation of Misspecified Models}. Econometrica, 50, 1--25.
\end{thebibliography}
\end{document}